\documentclass[letterpaper,twocolumn,10pt]{article}

\usepackage{titlesec}

\usepackage[hyphens]{url}
\usepackage{hyperref}
\usepackage{usenix2019_v3}

\usepackage[numbers,square,sort&compress]{natbib}
\usepackage{balance}
\usepackage[font=small,labelfont=bf]{caption}

\usepackage[pdftex]{graphicx}
\usepackage{grffile} \graphicspath{{figures/}}
\DeclareGraphicsExtensions{.pdf}

\usepackage[size=footnotesize,subrefformat=parens]{subcaption}

\usepackage[cmex10]{amsmath}

\usepackage{rotating}
\usepackage{multirow}
\usepackage{multicol}
\usepackage{makecell}
\usepackage{tabularx}
\usepackage{float} \newcolumntype{R}{@{}c}
\newcolumntype{D}{R@{\,\begin{rotate}{45}\rule[-.2ex]{11em}{0.5pt}\end{rotate}\,}}
\newcolumntype{S}{R@{\,\begin{rotate}{45}\rule[-.2ex]{8em}{0.5pt}\end{rotate}\,}}
\usepackage{booktabs}
\usepackage[flushleft]{threeparttable}

\usepackage{tikz}

\usepackage{cleveref}

\usepackage{versions}

\usepackage{enumitem}
\setitemize{noitemsep,topsep=0pt,parsep=0pt,partopsep=0pt}

\newlength\bubblesize
\newcommand{\yes}{
  \begin{tikzpicture}[baseline=0, line width=0.2ex]    \path[use as bounding box]  (\bubblesize,\bubblesize) circle (\bubblesize-.5\pgflinewidth);    \fill[black] (\bubblesize,\bubblesize) circle (\bubblesize);
  \end{tikzpicture}}
\newcommand{\no}{-}

\PassOptionsToPackage{hyphens}{url}
\makeatletter
\g@addto@macro{\UrlBreaks}{\UrlOrds}
\makeatother

\newcommand{\etal}{{\em et al.~}}

\newcommand{\singlefigwidthfactor}{0.8}

\renewcommand{\paragraph}[1]{\noindent\textbf{#1:}}

\newcommand\load{\ensuremath{\ell}}

\newcommand{\papertitle}{Once is Never Enough:\\ Foundations for Sound Statistical Inference in Tor Network Experimentation}

\hypersetup{
     bookmarks=true,
     colorlinks=true,
     linkcolor=black,
     citecolor=black,
     filecolor=black,
     urlcolor=black,
     pdfauthor = {Rob Jansen, Justin Tracey, and Ian Goldberg},      pdftitle = {\papertitle},
     pdfkeywords = {Tor, security, anonymity, performance, measurement}
}

\titlespacing\section{0pt}{6pt plus 4pt minus 2pt}{6pt plus 2pt minus 2pt}
\titlespacing\subsection{0pt}{6pt plus 4pt minus 2pt}{6pt plus 2pt minus 2pt}
\titlespacing\subsubsection{0pt}{6pt plus 4pt minus 2pt}{6pt plus 2pt minus 2pt}

\begin{document}

\title{\papertitle}

\date{}

\author{
{\rm Rob Jansen}\\
{U.S. Naval Research Laboratory} \\
{\rm rob.g.jansen@nrl.navy.mil}
\and
{\rm Justin Tracey}\\
{University of Waterloo}\\
{\rm j3tracey@uwaterloo.ca}
\and
{\rm Ian Goldberg}\\
{University of Waterloo}\\
{\rm iang@uwaterloo.ca}
} 
\maketitle

\begin{abstract}
Tor is a popular low-latency anonymous communication system that
focuses on usability and performance: a faster network will
attract more users, which in turn will improve the anonymity of everyone using
the system. The standard practice for previous research attempting to enhance Tor
performance is to draw conclusions from the observed results of a \textit{single} simulation for
standard Tor and for each research variant.
But because the simulations are run in \textit{sampled} Tor networks, it is possible that 
sampling error alone could cause the observed effects.
Therefore, we call into question the practical meaning of any conclusions that are drawn without considering
the statistical significance of the reported results.

In this paper, we build foundations upon which we 
improve the Tor experimental method.
First, we present a new Tor network modeling methodology
that produces \textit{more representative} Tor networks as well as new and improved experimentation tools
that run Tor simulations \textit{faster} and at a \textit{larger scale} than was
previously possible. We showcase these contributions by running simulations with
6,489 relays and 792k
simultaneously active users, the largest known Tor network simulations and the
first at a network scale of 100\%. 
Second, we present new statistical methodologies through which we: (i) show that running
\textit{multiple} simulations in \textit{independently sampled}
networks is
necessary in order to produce informative results; and (ii) show how to use the
results from multiple simulations to conduct sound \textit{statistical inference}.
We present a case study using 420~simulations to demonstrate how to apply our
methodologies to a concrete set of Tor experiments
and how to analyze the results.
\end{abstract}

\vspace{-2mm}
\section{Introduction} \label{sec:intro}
\vspace{-2mm}

Tor~\cite{tor-design} is a privacy-enhancing technology and the most
popular anonymous communication system ever deployed. Tor consists of
a network of relays that forward traffic on behalf of Tor users (i.e.,
clients) and Internet destinations. The Tor Project estimates that
there are about 2M daily active Tor
users~\cite{loesing2010measuring}, while recent privacy-preserving
measurement studies estimate that there are about 8M daily active
users~\cite{torusage-imc2018} and 792k simultaneously active users~\cite{tmodel-ccs2018}.
Tor is used for a variety of reasons, including
blocking trackers, defending against surveillance, resisting
fingerprinting and censorship, and freely browsing the
Internet~\cite{torproject}.

The usability of the Tor network is fundamental to the security it can
provide~\cite{usability:weis2006}; 
prior work has shown that real-world adversaries
\textit{intentionally} degrade usability to cause users to switch to less
secure communication protocols~\cite{Aryan2013a}.
Good usability enables Tor to
retain more users~\cite{privatelywaiting}, and more users generally
corresponds to better anonymity~\cite{econymics}.
Tor has made improvements in three primary usability components:
(i) the design of
the interface used to access and use the network (i.e., Tor Browser)
has been improved through usability 
studies~\cite{clark2007usability,norcie2012stoppoints,lee2017usability};
(ii) the performance perceived by Tor users has improved through
the deployment of new traffic scheduling 
algorithms~\cite{kist-tops2018,ccs10-scheduling};
and (iii) the network resources available for forwarding
traffic has grown from about 100\,Gbit/s to about 400\,Gbit/s in the
last 5 years~\cite{tormetrics}.
Although these changes have contributed to user growth, 
continued growth in the Tor network is desirable---not only because user growth
improves anonymity~\cite{econymics}, but also because access to
information is a universal and human right~\cite{freeinfo} and growth
in Tor means more humans can safely, securely, privately, and freely
access information online.

Researchers have contributed numerous proposals for improving Tor performance in
order to support continued growth in the network, including those that attempt
to improve Tor's path
selection~\cite{tor-relayselection,johnson2017avoiding,barton2016denasa,alsabah2013path,lin2016scalable,icoin15,rochet2017waterfilling,yang2015enhancing,shirazi2015towards,hanley2019dpselect,mitseva2020security,imani2019modified,rochet2020claps},
load balancing~\cite{jansen2010recruiting,Moore2011tortoise,jansen2013lira,geddes2016abra,johnson2017peerflow,imani2018guard},
traffic admission control~\cite{AlSabah2013pctcp,jansen2014kist,kist-tops2018,geddes2014imux,gopal2012torchestra,jansen2012throttling,mtor-cns15,Liu2017a,dinh2020scaling,kiran2019anonymity},
and congestion control mechanisms~\cite{alsabah2011defenestrator,geddes2013low}.
The standard practice when proposing a new mechanism for Tor is to run a
\textit{single experiment} with each recommended configuration of the mechanism
and a \textit{single experiment} with standard Tor. Measurements of a
performance metric (e.g., download time) are taken during each experiment, the
empirical distributions over which are directly compared across experiments.
Unfortunately, the experiments (typically simulations or
emulations~\cite{torexptools}) are done in scaled-down Tor test networks that
are sampled from the state of the true network at a static point in
time~\cite{jansen2012cset}; only a \textit{single sample} is considered even
though in reality the network changes over time in ways that could change the
conclusions.
Moreover, statistical inference techniques (e.g., repeated trials and interval
estimates) are generally not applied during the analysis of results, leading to
questionable conclusions.
Perhaps due in part to undependable results, only a few Tor performance research
proposals have been deployed over the
years~\cite{kist-tops2018,ccs10-scheduling}
despite the abundance of available
research.

\paragraph{Contributions}
We advance the state of the art by building foundations for
conducting sound Tor performance research in two major ways:
(i)~we design and validate Tor experimentation \textit{models} and develop new
and improved modeling and experimentation \textit{tools} that together allow us
to create and run more representative Tor test networks faster than was
previously possible; and
(ii)~we develop \textit{statistical methodologies} that enable sound statistical
inference of experimentation results and demonstrate how to apply our
methodologies through a case study on a concrete set of Tor experiments.

\paragraph{Models and Tools}
In \S\ref{sec:netmodel} we present a new Tor network modeling methodology that
produces \textit{more representative} Tor networks by considering the state of
the network \textit{over time} rather than at a static point as was previously
standard~\cite{jansen2012cset}.
We designed our modeling methodology to support the flexible generation of Tor
network models with configurable network, user, traffic load, and process scale
factors, supporting experiments in computing facilities with a range of
available resources. We designed our modeling tools such that expensive data
processing tasks need only occur once, and the result can be distributed to the
Tor community and used to efficiently generate any number of network models.

In \S\ref{sec:platforms} we contribute new and improved experimentation tools
that we optimized to enable us to run Tor experiments \textit{faster} and at a
\textit{larger scale} than was previously possible. In particular, we describe
several improvements we made to Shadow~\cite{jansen2012shadow}, the most popular
and validated platform for Tor experimentation, and demonstrate how our Tor
network models and improvements to Shadow increase the scalability of simulations. We showcase these contributions by running the largest known Tor
simulations---full-scale Tor networks with 6,489 relays and 792k simultaneously active users. We also
run smaller-scale networks of 2,000 relays and 244k users to compare to prior
work: we observe a reduction in RAM usage of 1.7\,TiB
(64\%) and a reduction in
run time of 33\,days, 12\,hours (94\%) compared to the state of the
art~\cite{tmodel-ccs2018}.

\paragraph{Statistical Methodologies}
In \S\ref{sec:significance} we describe a methodology that enables us to conduct
sound statistical inference using the results collected from scaled-down
(sampled) Tor networks. We find that running \textit{multiple} simulations in
\textit{independently sampled}
networks is necessary in order to
obtain statistically significant results, a methodology that has never before
been implemented in Tor performance research and causes us to question the
conclusions drawn in previous work (see \S\ref{sec:related:perf}).
We describe how to use multiple networks to estimate the distribution of a random
variable and compute confidence intervals over that distribution, and discuss
how network sampling choices would affect the estimation.

In \S\ref{sec:casestudy} we present a case study in order to demonstrate how to apply our modeling and
statistical methodologies to conduct sound Tor performance research. We present
the results from a total of 420 Tor simulations across three network scale and two
traffic load factors.
We find that the precision of the conclusions that can be drawn from the
networks used for simulations are dependent upon the scale of those networks.
Although it is possible to derive similar conclusions from networks of
different scales, fewer simulations are generally required in larger-scale than
smaller-scale networks to achieve a similar precision.
We conclude that one simulation is never enough to achieve statistically significant results.

\paragraph{Availability}
Through this work we have developed new modeling tools and improvements to
Shadow that we have released as open-source software as part of OnionTrace~v1.0.0,
TorNetTools~v1.1.0, TGen~v1.0.0, and
Shadow~v1.13.2.\footnote{https://github.com/shadow/\{\href{https://github.com/shadow/oniontrace}{oniontrace},\href{https://github.com/shadow/tornettools}{tornettools},\href{https://github.com/shadow/tgen}{tgen},\href{https://github.com/shadow/shadow}{shadow}\}}
We have made these and other research artifacts publicly available.\footnote{\url{https://neverenough-sec2021.github.io}}

\section{Background and Related Work} \label{sec:related}
\vspace{-1mm}
We provide a brief background on Tor before describing prior work on Tor
experimentation, modeling, and performance.

\subsection{Tor}

A primary function of the Tor network is to anonymize user traffic~\cite{tor-design}. To
accomplish this, the Tor network is composed of a set of Tor
\textit{relays} that forward traffic through the network on behalf of
\textit{users} running
Tor \textit{clients}. Some of the relays serve as \textit{directory
authorities} and are responsible for publishing a \textit{network consensus}
document containing relay information that is required to connect to and
use the network (e.g., addresses, ports, and fingerprints of
cryptographic identity keys for all relays in the network). Consensus documents also
contain a \textit{weight} for each relay to support a weighted path
selection process that attempts to balance traffic load across relays
according to relay bandwidth capacity.
To use the network, clients build long-lived \textit{circuits} through
a telescoping path of relays: the first in the path is called the \textit{guard}
(i.e., entry), the last is called the \textit{exit}, and the remaining
are called \textit{middle} relays. Once a circuit is established, the
client sends commands through the circuit to the exit
instructing it to open \textit{streams} to Internet
destinations (e.g., web servers); the request and response traffic for
these streams are multiplexed over the same circuit.
Another, less frequently used function of the network is to support
\textit{onion services} (i.e., anonymized servers) to which Tor
clients can connect (anonymizing both the client and the onion
service to the other).

\subsection{Tor Experimentation Tools}

Early Tor experimentation tools included packet-level simulators that
were designed to better understand the effects of Tor incentive
schemes~\cite{jansen2010recruiting,incentives-fc10}. Although these
simulators reproduced some of Tor's logic, they did not actually
use Tor source code and quickly became outdated and
unmaintained. Recognizing the need for a more realistic Tor
experimentation tool, researchers began developing tools following two
main approaches: network emulation and network simulation~\cite{torexptools}.

\looseness-1
\paragraph{Network Emulation}
ExperimenTor~\cite{bauer2011experimentor} is a Tor experimentation
testbed built on top of the ModelNet~\cite{modelnet} network emulation
platform. ExperimenTor consists of two components that generally run
on independent servers (or clusters): one component runs client
processes and the other runs the ModelNet core emulator that connects
the processes in a virtual network topology. The performance of this
architecture was improved in SNEAC~\cite{singh2014thesis} through the
use of Linux Containers and the kernel's network emulation module
\texttt{netem}, while tooling and
network orchestration were improved in NetMirage~\cite{netmirage}.

\paragraph{Network Simulation}
Shadow~\cite{jansen2012shadow} is a hybrid
discrete-event network simulator that runs applications as plugins.
We provide more background on Shadow in \S\ref{sec:platforms:background}.
Shadow's original design was improved with the addition of a
user-space non-preemptive thread scheduler~\cite{miller2015shadow}, and
later with a high performance dynamic loader~\cite{tracey2018elfs}.
Additional contributions have been made through several research
projects~\cite{jansen2014kist,kist-tops2018,tmodel-ccs2018},
and we make further contributions that improve Shadow's efficiency and
correctness as described in \S\ref{sec:platforms:improve}.

\subsection{Tor Modeling}
\vspace{-2mm}
An early approach to model the Tor network was developed for both Shadow and
ExperimenTor~\cite{jansen2012cset}.
The modeling approach 
produced scaled-down Tor test networks by sampling relays and their attributes
from a \textit{single} true Tor network consensus.
As a result, the models
are particularly sensitive to short-term
temporal changes in the composition of the true network  (e.g., those that
result from natural relay churn, network attacks, or misconfigurations).
The new techniques we present in
\S\ref{sec:tormodel} are more robust to such variation because they are
designed to use Tor metrics data spanning a user-selectable time period 
(i.e., from any chosen set of consensus files) 
in order to create simulated Tor networks that are more
\textit{representative} of the true Tor network over time.

In previous models, the number of clients to use and their behavior
profiles were unknown, so finding a suitable combination of traffic
generation parameters that would yield an appropriate amount of
background traffic was often a challenging and iterative process. But
with the introduction of privacy-preserving measurement
tools~\cite{PrivExElahi2014,privcount-ccs2016,mani2017historvarepsilon,fenske2017distributed}
and the recent publication of Tor measurement 
studies~\cite{privcount-ccs2016,tmodel-ccs2018,torusage-imc2018}, 
we have gained a more informed understanding of the traffic
characteristics of Tor. Our new modeling techniques use
Markov models informed by (privacy-preserving) statistics
from true Tor traffic~\cite{tmodel-ccs2018}, while significantly
improving experiment scalability as we demonstrate in
\S\ref{sec:platforms:validate}.

\subsection{Tor Performance Studies}
\label{sec:related:perf}

The Tor experimentation tools and models described above have assisted
researchers in exploring how changes to Tor's
path selection~\cite{tor-relayselection,johnson2017avoiding,barton2016denasa,alsabah2013path,lin2016scalable,icoin15,rochet2017waterfilling,yang2015enhancing,shirazi2015towards,hanley2019dpselect,mitseva2020security,imani2019modified,rochet2020claps},
load balancing~\cite{jansen2010recruiting,Moore2011tortoise,jansen2013lira,geddes2016abra,johnson2017peerflow,imani2018guard},
traffic admission control~\cite{AlSabah2013pctcp,jansen2014kist,kist-tops2018,geddes2014imux,gopal2012torchestra,jansen2012throttling,mtor-cns15,Liu2017a,dinh2020scaling,kiran2019anonymity},
congestion control~\cite{alsabah2011defenestrator,geddes2013low},
and denial of service mechanisms~\cite{jansen2014sniper,hopper2014challenges,conrad2014analyzing,rochet2018dropping,jansen2019point}
affect Tor performance and security~\cite{alsabah2016performance}.
The standard practice that has emerged from this work is to sample a single
scaled-down Tor network model and use it to run experiments with standard Tor
and each of a set of chosen configurations of the proposed performance-enhancing
mechanism. Descriptive statistics or empirical distributions of the results are
then compared across these experiments.
Although some studies use multiple trials of each experimental configuration
in the \textit{chosen} sampled
network~\cite{jansen2013lira,johnson2017peerflow}, none of them involve running
experiments in \textit{multiple} sampled networks, which is necessary to estimate
effects on the real-world Tor network (see \S\ref{sec:significance}).
Additionally, statistical inference techniques (e.g., interval estimates) are not
applied during the analysis of the results, leading to questions about the
extent to which the conclusions drawn in previous work are relevant to the real
world.
Our work advances the state of the art of the experimental process
for Tor performance research: in \S\ref{sec:significance} we describe
new statistical methodologies that enable researchers to conduct sound
statistical inference from Tor experimentation results, and in
\S\ref{sec:casestudy} we present a case study to demonstrate how to put our
methods into practice.

\section{Models for Tor Experimentation} \label{sec:netmodel}

In order to conduct Tor experiments that produce meaningful results,
we must have network and traffic models that accurately represent the
composition and traffic characteristics of the Tor network.
In this section, we describe new modeling techniques that make use of the
latest data from recent privacy-preserving measurement
studies~\cite{tmodel-ccs2018,torusage-imc2018,privcount-ccs2016}.
Note that while exploring alternatives for every modeling choice that will be described
in this section is out of scope for this paper, we will discuss some alternatives that are
worth considering in \S\ref{sec:conc}.

\subsection{Internet Model}
\label{sec:inetmodel}

Network communication is vital to distributed systems; the bandwidth and the network
latency
between nodes are primary characteristics that affect
performance. Jansen \etal have produced an Internet
map~\cite{tmodel-ccs2018} that we find useful for our
purposes; we briefly describe how it was constructed before
explaining how we modify it.

To produce an Internet map, Jansen \etal\cite{tmodel-ccs2018} conducted Internet
measurements using globally distributed vantage points (called \textit{probes}) from the
RIPE Atlas measurement system (\href{https://atlas.ripe.net}{atlas.ripe.net}). They assigned a
representative probe for each of the 1,813 cities in which at least
one probe was available. They used \texttt{ping} to estimate the
latency between all of the 1,642,578 distinct pairs of representative
probes, and they crawled \href{https://www.speedtest.net}{speedtest.net} to extract
upstream and downstream bandwidths for each city.\footnote{\href{https://www.speedtest.net}{speedtest.net} ranks mobile and fixed broadband speeds around the world.}
They encoded the
results into an Internet map stored in the \texttt{graphml} file
format; each vertex corresponds to a representative probe and encodes
the bandwidth available in that city, and each edge corresponds to a
path between a pair of representative probes and encodes the network
latency between the pair.

Also encoded on edges in the Internet map were packet loss rates. Each
edge $e$ was assigned a packet loss rate $p_e$ according to the
formula $p_e \gets 0.015 \cdot L_e / 300$ where $L_e$ is the latency
of edge $e$. This improvised formula was not based on any real data.
Our experimentation platform (described in
\S\ref{sec:platforms}) already includes for each host an edge router
component that drops packets when buffers are full. Because
additional packet loss from core routers is uncommon~\cite{corerouters},
we modify
the Internet map by setting $p_e$ to zero for all edges.\footnote{Future work
	should consider developing a more realistic packet loss model that is, e.g., based on
	measurements of actual Tor clients and relays.}
We use the resulting Internet model in all simulations in this paper.

\subsection{Tor Network Model} \label{sec:tormodel}

To the Internet model we add hosts that run Tor relays and form a Tor
overlay network. The Tor modeling task is to choose host bandwidths,
Internet locations, and relay configurations that support the creation
of Tor test networks that are representative of the true Tor
network.

We construct Tor network models in two phases: \textit{staging} and
\textit{generation}. The two-phase process allows us to perform the
computationally expensive staging phase once, and then perform the
computationally inexpensive generation phase any number of times. It
also allows us to release the staging files to the community, whose
members may then use our Tor modeling tools without first
processing large datasets.

\begin{table}[t]
\centering
\footnotesize
\captionsetup{skip=2pt} \begin{threeparttable}
\caption{Statistics computed during the \textit{staging} phase.}
\label{tab:stage}
\begin{tabular}{cl}
\toprule
\textbf{Stat.} & \textbf{Description} \\
\midrule
  $r_i$ & the fraction of consensuses in which relay $i$ was running\\
  $g_i$ & the fraction of consensuses in which relay $i$ was a guard\\
  $e_i$ & the fraction of consensuses in which relay $i$ was an exit\\
  $w_i$ & the median normalized consensus weight of relay $i$\\
  $b_i$ & the max observed bandwidth of relay $i$\\
  $\lambda_i$ & the median bandwidth rate of relay $i$\\
  $\beta_i$ & the median bandwidth burst of relay $i$\\
\midrule
  $C_{\rho}$ & median across consensuses of relay count for each position $\rho^{\dagger}$\\
  $W_{\rho}$ & median across consensuses of total weight for position $\rho^{\dagger}$\\
\midrule
  $U_c$ & median normalized probability that a user is in country $c^{\ddagger}$\\
\bottomrule
\end{tabular}
\begin{tablenotes}
\item[$\dagger$] Valid positions are $D$: exit+guard, $E$: exit, $G$: guard, and $M$: middle.
\item[$\ddagger$] Valid countries are any two-letter country code (e.g., $us$, $ca$, etc.).
\end{tablenotes}
\end{threeparttable}
\end{table}

\subsubsection{Staging} \label{sec:tormodel:stage}
Ground truth details about the temporal composition and state of the Tor network
are available in Tor network data files (i.e., hourly network consensus and
daily relay server descriptor files) which have been published since 2007.
We first gather the subset of these files that
represents the time period that we want to model (e.g., all files published in
January 2019), and then extract network attributes
from the files in the staging
phase so that we can make use of them in the networks we later generate.
In addition to
extracting the IP address, country code, and fingerprint
of each
relay~$i$, we compute the per-relay and network summary
statistics shown in Table~\ref{tab:stage}.
We also process the Tor \textit{users} dataset
containing per-country user counts, which Tor has published daily
since 2011~\cite{tormetrics}. From this data we compute the median normalized
probability that a user appears in each country. We store the results
of the staging phase in two small \texttt{JSON} files (a few MiB
each) that we use in the generation phase.
Note that we could make use of other
network information if it were able to be safely measured and published
(see Appendix~\ref{sec:ontol} for an ontology of some independent variables that could be useful).

\subsubsection{Generation} \label{sec:tormodel:generate}
In the generation phase, we use the data extracted during the staging
phase
and the results from a recent privacy-preserving Tor measurement
study~\cite{tmodel-ccs2018} to generate Tor network models of a
configurable \textit{scale}. For example, a \textit{100\% Tor network}
represents a model of equal scale to the true Tor network. Each
generated model is stored in an \texttt{XML} \textit{configuration
file}, which specifies the hosts that should be instantiated,
their bandwidth properties and locations in the Internet map, the
processes they run, and configuration options for each process.
Instantiating a model will result in a Tor test network that
is representative of the true Tor network. We describe the
generation of the configuration file by the type of hosts that make up
the model: \textit{Tor network relays}, \textit{traffic generation},
and \textit{performance benchmarking}.

\paragraph{Tor Network Relays}
The relay staging file may contain more relays than we need for a 100\%
Tor network (due to relay churn in the network during the staged time period),
so we first choose enough relays for a 100\% Tor network by
sampling $n \gets \sum_{\rho} C_{\rho}$ relays without replacement, using each
relay's running frequency $r_i$ as its sampling weight.\footnote{Alternatives to
	weighted sampling should be considered if staging time periods during which the Tor
	network composition is extremely variable.}
We then assign
the guard and exit flag to each of the $n$ sampled relays~$j$ with a
probability equal to the fraction of consensuses in which relay~$j$
served as a guard $g_j$ and exit $e_j$, respectively.

To create a network whose scale is $ 0 < s \leq 1$ times the size of
the 100\% network,\footnote{Because of the RAM and CPU requirements (see
	\S\ref{sec:platforms}), we expect that it will generally be infeasible to run
	100\% Tor networks. The configurable scale $s$ allows for tuning the amount of resources required to run a model.}
we further subsample from the sampled set of $n$
relays to use in our scaled-down network model. We describe our
subsampling procedure for middle relays for ease of exposition, but
the same procedure is repeated for the remaining positions (see
Table~\ref{tab:stage} note$^{\dagger}$). To subsample $m \gets s \cdot C_M$
middle relays, we: (i)\,sort the list of sampled middle relays by their
normalized consensus weight $w_j$, (ii)\,split the list into $m$
buckets, each of which contains as close as possible to an
equal number of relays,
and
(iii)\,from each bucket, select the relay with the median weight $w_j$
among those in the bucket. This strategy guarantees that the weight
distribution across relays in our subsample is a best fit to the weight distribution of relays in the
original sample~\cite{jansen2012cset}.

A new host is added to the configuration file for each subsampled
relay $k$. Each host is assigned the IP address and country code
recorded in the staging file for relay $k$, which will allow it to be
placed in the nearest city in the Internet map. The host running relay
$k$ is also assigned a symmetric bandwidth capacity equal to $b_k$;
i.e., we use the maximum observed bandwidth as our best estimate of a
relay's bandwidth capacity. Each host is configured to run a Tor relay
process that will receive the exit and guard flags that we assigned
(as previously discussed), and each relay $k$ sets its token bucket
rate and burst options to $\lambda_k$ and $\beta_k$, respectively.
When executed, the relay processes will form a functional Tor
overlay network capable of forwarding user traffic.

\paragraph{Traffic Generation}
A primary function of the Tor network is to forward traffic on behalf
of users.
To accurately characterize Tor network usage, we use the following
measurements from a recent privacy-preserving Tor measurement
study~\cite{tmodel-ccs2018}: the total number of active users $\phi=792$k
(counted at guards) and the total number of active circuits $\psi=1.49$M
(counted at exits) in an average 10 minute period.

To generate a Tor network whose scale is $0 < s \leq 1$ times the size
of the 100\% network, we compute the total number of users we need to
model as $u \gets s \cdot \phi$.
We compute the total number of circuits that those $u$ users create every 10
minutes as $c \gets \load \cdot s \cdot \psi$, where  $\load \geq 0$ is a load
factor that allows for configuration of the amount of traffic load generated by
the $u$ users ($\load = 1$ results in ``normal'' traffic load).
We use a process scale factor
$0 < p
\le 1$ to allow for configuration of the number of Tor client
processes that will be used to generate traffic on the $c$ circuits
from the $u$ users.
Each of $p \cdot u$ Tor client processes will support
the combined traffic of $1/p$ users, i.e., the traffic from $\tau
\gets c/p \cdot u$ circuits.

The $p$ factor can be used to significantly reduce the amount of RAM and CPU
resources required to run our Tor model; e.g., setting $p=0.5$ means we only
need to run half as many Tor client processes as the number of users we are
simulating.\footnote{A primary effect of $p<1$ is fewer network descriptor
	fetches, the network impact of which is negligible relative to the total
	traffic generated.}
At the same time, $p$ is a reciprocal factor w.r.t. the traffic
that each Tor client generates; e.g., setting $p=0.5$ causes each client to
produce twice as many circuits (and the associated traffic) as a single user would.

We add $p \cdot u$ new traffic generation client hosts to our configuration
file. For each such client, we choose a country according to the
probability distribution $U$, and assign the client to a random city in that
country using the Internet map in \S\ref{sec:inetmodel}.\footnote{Shadow will
	arbitrarily choose an IP address for the host such that it can route packets to
	all other simulation hosts (clients, relays, and servers).}
Each client runs a Tor process in
client mode configured to disable guards\footnote{Although a Tor client uses
	guards by default, for us it would lead to inaccurate load balancing because
	each client simulates $1/p$ users. Support in the Tor client for running
	multiple ($1/p$) parallel guard ``sessions'' (i.e., assigning a guard to each
	user ``session'') is an opportunity for future work.}
and a TGen traffic generation process that is configured to send its traffic
through the Tor client process over localhost (we significantly extend a
previous version of TGen~\cite[\S5.1]{tmodel-ccs2018} to support our models).
Each TGen process is configured to generate traffic using Markov models (as we
describe below), and we assign each host a bandwidth capacity equal to the
maximum of $10/p$~Mbit/s and 1~Gbit/s to prevent it from biasing the traffic
rates dictated by the Markov models when generating the combined traffic of
$1/p$ users. Server-side counterparts to the TGen processes are also added to
the configuration file (on independent hosts).

Each TGen process uses three Markov models to accurately model Tor traffic
characteristics: (i)\;a \textit{circuit} model, which captures the circuit
inter-arrival process on a per-user basis; (ii)\;a \textit{stream} model, which
captures the stream inter-arrival process on a per-circuit basis; and (iii)\;a
\textit{packet} model, which captures the packet inter-arrival process on a
per-stream basis.
Each of these models are based on a recent privacy-preserving measurement study
that used PrivCount~\cite{privcount-ccs2016} to collect measurements of real
traffic being forwarded by a set of Tor exit relays~\cite{tmodel-ccs2018}.
We encode the \textit{circuit} inter-arrival process as a simple single state
Markov model that emits new circuit events according to an exponential
distribution with rate $1/\mu/\tau$ microseconds, where $\mu \gets 6 \cdot 10^8$
is the number of microseconds in 10 minutes. New streams on each circuit and
packets on each stream are generated using the \textit{stream} and
\textit{packet} Markov models, respectively, which were directly measured in Tor
and published in previous work~\cite[\S5.2.3]{tmodel-ccs2018}.

The rates and patterns of the traffic generated using the Markov models will
mimic the rates and patterns of real Tor users: the models encode common
distributions (e.g., exponential and log-normal) and their parameters, such that
they can be queried to determine the amount of time to wait between the creation
of new circuits and streams and the transfer of packets (in both the send and
receive directions).

Each TGen client uses unique seeds for all Markov models so that it generates
unique traffic characteristics.\footnote{The Markov model seeds are unique
	across clients, but generated from the same master seed in order to maintain a
	deterministic simulation.} Each TGen client also creates a unique \texttt{SOCKS}
username and password for each generated circuit and uses it for all Tor streams
generated in the circuit; due to Tor's \texttt{IsolateSOCKSAuth} feature, this
ensures that streams from different circuits will in fact be assigned to
independent circuits.

We highlight that although prior work also made use of the \textit{stream} and \textit{packet} Markov
models~\cite[\S5.2.3]{tmodel-ccs2018}, we extend previous work with
a \textit{circuit} Markov model that can be used to
continuously generate circuits independent of the length of an
experiment. Moreover, previous work did not
consider either load scale $\load$ or process scale $p$;
$\load$ allows for research under varying levels of congestion,
and our optimization of simulating $1/p$ users in each Tor client process
allows us to more quickly run significantly
larger network models than we otherwise could (as we will show in
\S\ref{sec:platforms:validate}).

\paragraph{Performance Benchmarking}
The Tor Project has published performance benchmarks since
2009~\cite{tormetrics}. The benchmark process downloads 50 KiB, 1 MiB,
and 5 MiB files through the Tor network several times per hour, and
records various statistics about each download including the time to
download the first and last byte of the files. We mirror this process
in our models; running several benchmarking clients that use some of the same code
as Tor's benchmarking clients (i.e., TGen)
allows us to
\textit{directly} compare the performance obtained in our simulated
Tor
networks with that of the true Tor network.

\subsubsection{Modeling Tools}
\label{sec:modeltools}

We implemented several tools that we believe are fundamental to our ability to
model and execute realistic Tor test networks. We have released these tools as
open source software to help facilitate Tor research:
(i)\,a new Tor network modeling toolkit called TorNetTools (3,034 LoC) that implements our
modeling algorithms from \S\ref{sec:tormodel:generate};
(ii)\,extensions and enhancements to the TGen traffic
generator~\cite[\S5.1]{tmodel-ccs2018} (6,531 LoC added/modified and
1,411 removed) to support our traffic generation models; and
(iii)\,a new tool called OnionTrace (2,594~LoC) to interact
with a Tor process and improve reproducibility of experiments.
We present additional details about these tools in
	Appendix~\ref{sec:netmodel:tools:appendix}.

\section{Tor Experimentation Platform} \label{sec:platforms}

The models that we described in \S\ref{sec:netmodel} could reasonably
be instantiated in a diverse set of experimentation platforms in order
to produce representative Tor test networks. We use
Shadow~\cite{jansen2012shadow}, the most popular and validated
platform for Tor experimentation. We provide a brief background on
Shadow's design, explain the improvements we made to support accurate
experimentation, and show how our improvements and models from
\S\ref{sec:netmodel} contribute to the state of the art.

\subsection{Shadow Background} \label{sec:platforms:background}

Shadow is a hybrid experimentation platform~\cite{jansen2012shadow}.
At its core, Shadow is a conservative-time discrete-event network
simulator: it simulates hosts, processes, threads, TCP and UDP,
routing, and other kernel operations.
One of Shadow's advantages is that it dynamically
loads real applications as plugins and directly executes them as
native code. In this regard, Shadow emulates a network and a Linux
environment: applications running as plugins should function as they
would if they were executed on a bare-metal Linux installation.

Because Shadow is a user-space, single process application, it can
easily run on laptops, desktops, and servers with minimal
configuration (resource requirements depend on the size of the
experiment model). As a simulator, Shadow has complete control over
simulated time; experiments may run faster or slower than real time
depending on: (i) the simulation load relative to the processing
resources available on the host machine, and (ii) the inherent
parallelizability of the experiment model. This control over time
decouples the fidelity of the experiment from the processing time
required to execute it, and allows Shadow to scale independently of
the processing capabilities of the host machine; Shadow is usually
limited by the RAM requirements of its loaded plugins.

Shadow has numerous features that allow it to achieve its goals,
including dynamic loading of independent namespaces for plugins~\cite{tracey2018elfs},
support for multi-threaded plugins via a non-preemptive concurrent
thread scheduling library (GNU Portable
Threads\footnote{\url{https://www.gnu.org/software/pth}})~\cite{miller2015shadow}, function
interposition, and an event scheduler based on work
stealing~\cite{worksteal}. The combination of its features makes
Shadow a powerful tool for Tor experimentation, and has led it to
become the most popular and standard tool for conducting Tor
performance research~\cite{torexptools}.

\subsection{Shadow Improvements} \label{sec:platforms:improve}

After investigation of the results from some early experiments, we made
several improvements to Shadow that we believe cause it to
produce significantly more accurate results when running our Tor
network models from \S\ref{sec:tormodel}.
Our improvements include run-time optimizations, fixes to ensure
deterministic execution, faster Tor network bootstrapping, more
realistic TCP connection limits, and several network stack
improvements (see
  Appendix~\ref{sec:platforms:improve:appendix} for more details).
Our improvements have been incorporated into Shadow~v1.13.2.

\subsection{Evaluation}
\label{sec:platforms:validate}
\newcommand{\metricshspace}{-2mm}

We have thus far made two types of foundational
contributions: those that result in more \textit{representative Tor networks}, and
those that allow us to run more \textit{scalable simulations} faster than was
previously possible. We demonstrate these contributions through Tor network
simulations in Shadow.

\begin{figure}[t]
	\captionsetup{skip=0pt} 	\centering
	\includegraphics[width=\singlefigwidthfactor\columnwidth]{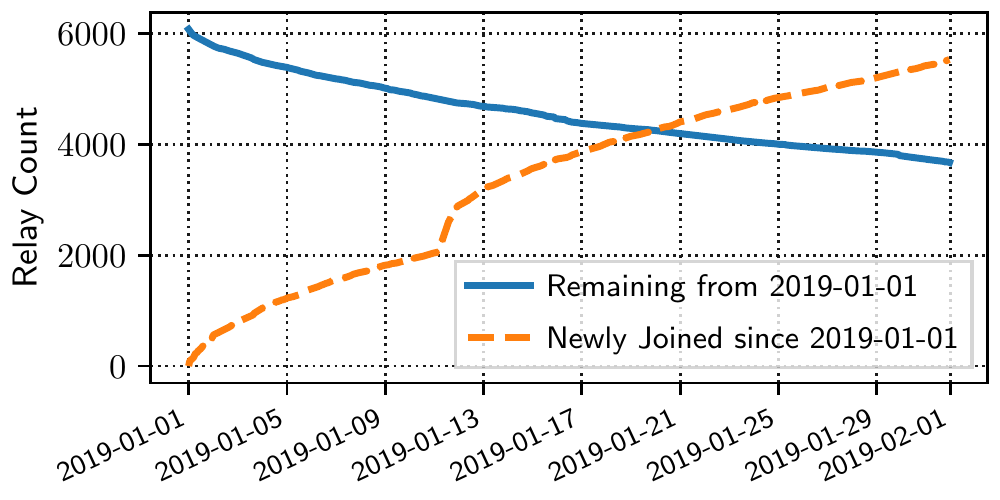}
	\caption{\hspace{-1mm}
		The rate of Tor relay churn over all 744 consensuses from January 2019. Shown
		are the number of Tor relays that existed on 2019-01-01 that remain and the number
		of relays that did not exist on 2019-01-01 that joined (and possibly left
		again) over time.
	}
	\label{fig:relay_churn}
\end{figure}

\paragraph{Representative Networks}
We produce more representative networks by considering the state of the network
over time rather than modeling a single snapshot as did previous
work~\cite{jansen2012cset, tmodel-ccs2018}. We consider relay churn to
demonstrate how the true Tor network changes over time.
Figure~\ref{fig:relay_churn} shows the rate of relay churn over all 744
consensus files (1 per hour) in Tor during January 2019. After 2 weeks,
fewer than 75\% of relays that were part of the network on 2019-01-01 still
remain while more than 3,000 new relays joined the network. After 3 weeks,
more new relays had joined the network than had remained since 2019-01-01. Our
models account for churn by sampling from \textit{all} such relays as described
in \S\ref{sec:tormodel}.

In addition to producing more representative models, our Shadow network stack
enhancements further improve network accuracy. To demonstrate these
contributions, we simulate ten Tor network models that were generated following
the methods in \S\ref{sec:tormodel} (using Tor network state from 2019-01).
We model Tor at the same $s=0.31$ scale that was used in previous
work~\cite{tmodel-ccs2018} (i.e., $\approx$2k relays and $\approx$250k users)
using a process scale factor of $p=0.01$ (i.e., each TGen process simulated
$1/0.01=100$ Tor users).
We compare our simulation results to those produced by state-of-the-art
methods~\cite{tmodel-ccs2018} (which used Tor network state from 2018-01) and to
reproducible Tor metrics~\cite{tormetrics,tormetrics-repro} from the
corresponding modeling years (2019 for our work, 2018 for the CCS 2018
work).\footnote{Although the models used Tor data spanning one month, we
	consider it reasonable to reflect the general state of Tor throughout the
	respective year.}

\begin{figure}[t]
	\centering
		\hspace{7mm}\includegraphics[width=0.9\columnwidth]{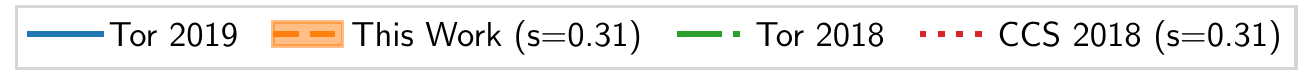}
	\\
	\includegraphics[width=3.5mm]{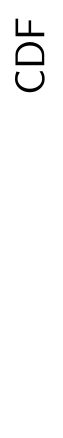}
	\hspace{\metricshspace}
	\begin{subfigure}[b]{0.2425\columnwidth}
		\centering
		\captionsetup{skip=0pt} 		\includegraphics[width=1.0\textwidth]{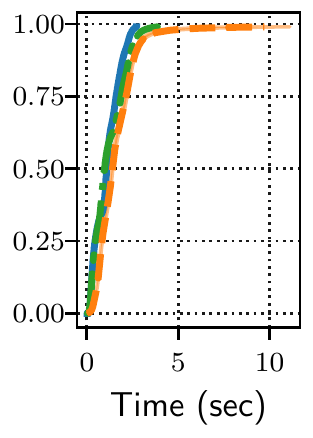}
		\caption{Circuit Build}
		\label{fig:tor:cbt}
	\end{subfigure}
	\hspace{\metricshspace}
	\begin{subfigure}[b]{0.2425\columnwidth}
		\centering
		\captionsetup{skip=0pt} 		\includegraphics[width=1.0\textwidth]{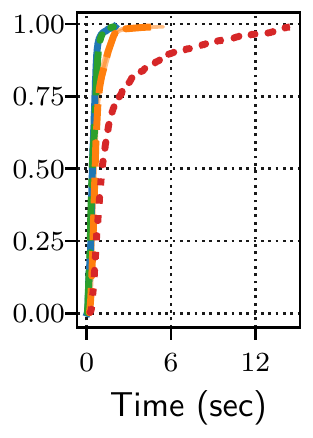}
		\caption{Circuit RTT}
		\label{fig:tor:crtt}
	\end{subfigure}
	\hspace{\metricshspace}
	\begin{subfigure}[b]{0.2425\columnwidth}
		\centering
		\captionsetup{skip=0pt} 		\includegraphics[width=1.0\textwidth]{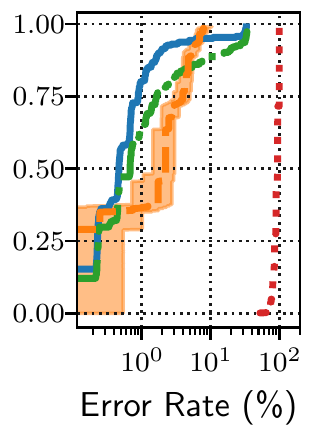}
		\caption{DL Error Rate}
		\label{fig:tor:clienterr}
	\end{subfigure}
	\hspace{\metricshspace}
	\begin{subfigure}[b]{0.2425\columnwidth}
		\centering
		\captionsetup{skip=0pt} 		\includegraphics[width=1.0\textwidth]{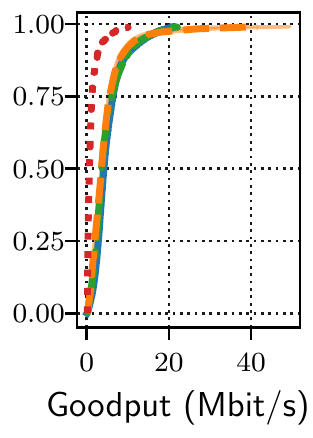}
		\caption{DL Goodput}
		\label{fig:tor:cgput}
	\end{subfigure}
	\\
	\includegraphics[width=3.5mm]{tor-compare-ylabel.pdf}
	\hspace{\metricshspace}
	\begin{subfigure}[b]{0.2425\columnwidth}
		\centering
		\captionsetup{skip=0pt,font=footnotesize} 		\includegraphics[width=1.0\textwidth]{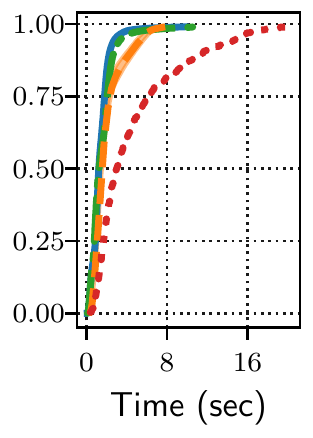}
		\caption{TTLB 50\,KiB}
		\label{fig:tor:ttlb50k}
	\end{subfigure}
	\hspace{\metricshspace}
	\begin{subfigure}[b]{0.2425\columnwidth}
		\centering
		\captionsetup{skip=0pt} 		\includegraphics[width=1.0\textwidth]{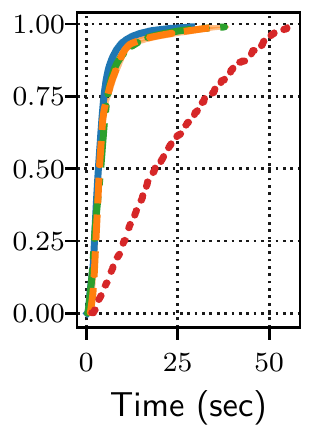}
		\caption{TTLB 1\,MiB}
		\label{fig:tor:ttlb1m}
	\end{subfigure}
	\hspace{\metricshspace}
	\begin{subfigure}[b]{0.2425\columnwidth}
		\centering
		\captionsetup{skip=0pt} 		\includegraphics[width=1.0\textwidth]{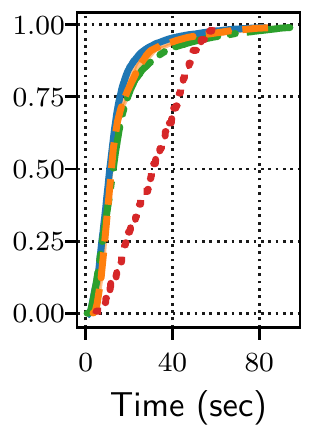}
		\caption{TTLB 5\,MiB}
		\label{fig:tor:ttlb5m}
	\end{subfigure}
	\hspace{\metricshspace}
	\begin{subfigure}[b]{0.2425\columnwidth}
		\centering
		\captionsetup{skip=0pt} 		\includegraphics[width=1.0\textwidth]{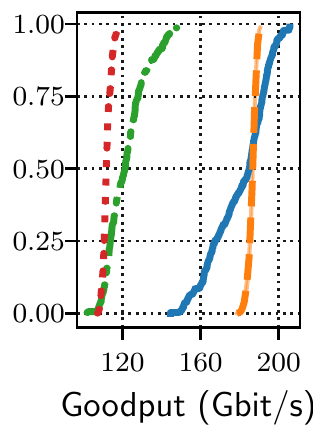}
		\caption{Relay Goodput}
		\label{fig:tor:netgput}
	\end{subfigure}
	\caption{
		Results from 10
		simulations at network scale $s=0.31$ (modeled using Tor network state from
		2019-01) and 1 simulation using state-of-the-art methods from CCS
		2018~\cite{tmodel-ccs2018} (modeled using Tor network state from 2018-01) 
		compared to reproducible Tor metrics~\cite{tormetrics-repro} during the respective years.
				Shown are benchmark client metrics for: 
		\subref{fig:tor:cbt}\,circuit build times;
		\subref{fig:tor:crtt}\,round trip times (time from data request to first byte of response);
		\subref{fig:tor:clienterr}\,download error rate;
		\subref{fig:tor:cgput}\,download goodput (i.e., transfer rate for range [0.5\,MiB, 1\,MiB] over 1\,MiB and 5\,MiB transfers), and
		\subref{fig:tor:ttlb50k}--\subref{fig:tor:ttlb5m}\,download times for transfers of size 50\,KiB, 1\,MiB, and 5\,MiB.
		Relay goodput in \subref{fig:tor:netgput}\,is, for each second, the sum over all relays of application bytes written (extrapolated by a $1/0.31$ factor to account for scale).
		(Note that circuit times in \subref{fig:tor:cbt} are unavailable in the CCS 2018 model~\cite{tmodel-ccs2018}.)
		The shaded areas represent 95\% confidence intervals (CIs) that were computed
		following our method from \S\ref{sec:significance}.
	}
	\label{fig:tor}
\end{figure}

The results in Figure~\ref{fig:tor} generally show that previous work is noticeably less accurate when
compared to Tor\;2018 than our work is compared to Tor\;2019. We notice
that previous work exhibited a high client
download error rate in Figure~\ref{fig:tor:clienterr}
and significantly longer download times in
Figures~\ref{fig:tor:ttlb50k}--\ref{fig:tor:ttlb5m} despite the network being
appropriately loaded as shown in Figure~\ref{fig:tor:netgput}.
We attribute these errors to the connection limit and network stack
limitations that were present in the CCS 2018 version of Shadow (the errors are not
present in this work due to our Shadow improvements from \S\ref{sec:platforms:improve}).
Also, we remark that the relay goodput in Figure~\ref{fig:tor:netgput}
exhibits more variance in Tor than in Shadow because the Tor data is being
aggregated over a longer time period (1 year for Tor vs. less than 1 hour for Shadow)
during which the Tor network composition is significantly changing
(see Figure~\ref{fig:relay_churn}).

\begin{figure}[t]
	\centering
		\hspace{7mm}\includegraphics[width=0.9\columnwidth]{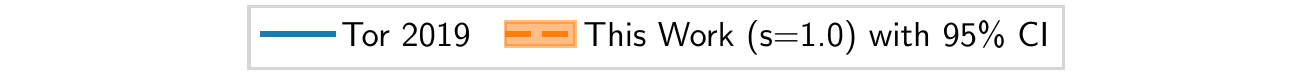}
	\\
	\includegraphics[width=3.5mm]{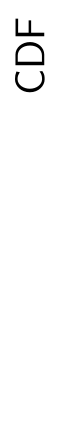}
	\hspace{\metricshspace}
	\begin{subfigure}[b]{0.2425\columnwidth}
		\centering
		\captionsetup{skip=0pt} 		\includegraphics[width=1.0\textwidth]{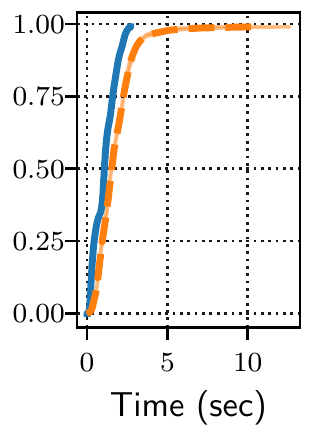}
		\caption{Circuit Build}
		\label{fig:torci:cbt}
	\end{subfigure}
	\hspace{\metricshspace}
	\begin{subfigure}[b]{0.2425\columnwidth}
		\centering
		\captionsetup{skip=0pt} 		\includegraphics[width=1.0\textwidth]{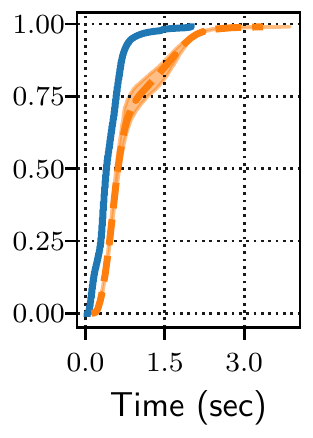}
		\caption{Circuit RTT}
		\label{fig:torci:crtt}
	\end{subfigure}
	\hspace{\metricshspace}
	\begin{subfigure}[b]{0.2425\columnwidth}
		\centering
		\captionsetup{skip=0pt} 		\includegraphics[width=1.0\textwidth]{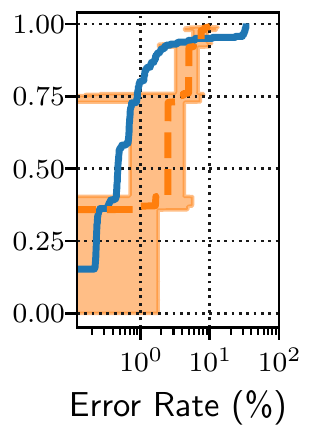}
		\caption{DL Error Rate}
		\label{fig:torci:clienterr}
	\end{subfigure}
	\hspace{\metricshspace}
	\begin{subfigure}[b]{0.2425\columnwidth}
		\centering
		\captionsetup{skip=0pt} 		\includegraphics[width=1.0\textwidth]{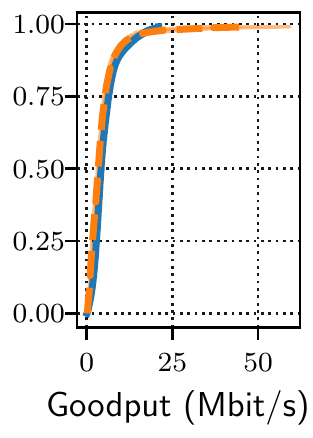}
		\caption{DL Goodput}
		\label{fig:torci:cgput}
	\end{subfigure}
	\\
	\includegraphics[width=3.5mm]{tor-compare-ylabel-ci.pdf}
	\hspace{\metricshspace}
	\begin{subfigure}[b]{0.2425\columnwidth}
		\centering
		\captionsetup{skip=0pt,font=footnotesize} 		\includegraphics[width=1.0\textwidth]{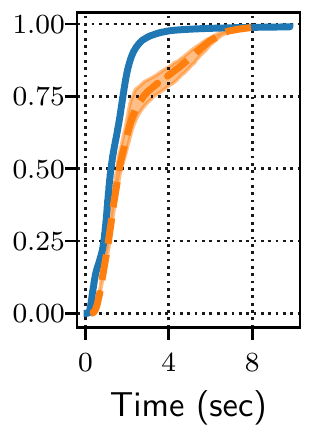}
		\caption{TTLB 50\,KiB}
		\label{fig:torci:ttlb50k}
	\end{subfigure}
	\hspace{\metricshspace}
	\begin{subfigure}[b]{0.2425\columnwidth}
		\centering
		\captionsetup{skip=0pt} 		\includegraphics[width=1.0\textwidth]{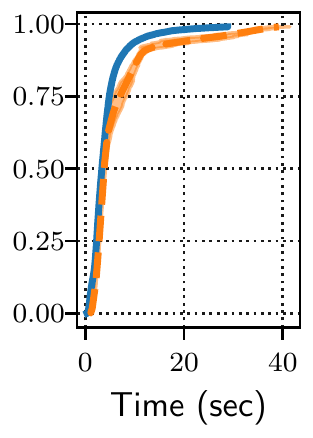}
		\caption{TTLB 1\,MiB}
		\label{fig:torci:ttlb1m}
	\end{subfigure}
	\hspace{\metricshspace}
	\begin{subfigure}[b]{0.2425\columnwidth}
		\centering
		\captionsetup{skip=0pt} 		\includegraphics[width=1.0\textwidth]{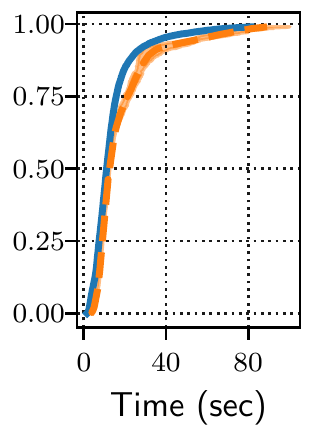}
		\caption{TTLB 5\,MiB}
		\label{fig:torci:ttlb5m}
	\end{subfigure}
	\hspace{\metricshspace}
	\begin{subfigure}[b]{0.2425\columnwidth}
		\centering
		\captionsetup{skip=0pt} 		\includegraphics[width=1.0\textwidth]{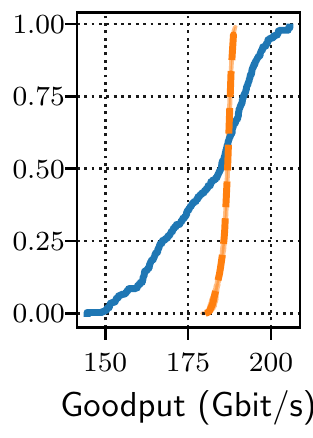}
		\caption{Relay Goodput}
		\label{fig:torci:netgput}
	\end{subfigure}
	\caption{
		Results from 3
		simulations at network scale $s=1.0$ (modeled using Tor
		network state from 2019-01) compared to reproducible Tor
		metrics~\cite{tormetrics-repro}.
		The metrics are as were defined in the Fig.~\ref{fig:tor} caption. The shaded areas
		represent 95\% confidence intervals (CIs) that were computed following our method
		from \S\ref{sec:significance}.
	}
	\label{fig:torci}
	\vspace{-2mm}
\end{figure}

\begin{table}[t]
	\centering
	\footnotesize
	\captionsetup{skip=2pt} 	\begin{threeparttable}
		\caption{Scalability improvements over the state of the art}
		\label{tab:scalability}
		\begin{tabular}{@{\,\,\,} c @{\,} c @{\,} c @{\,\,} c @{\,\,} c @{\,\,\,} c @{\,\,\,}}
			\toprule
			\textbf{Model} & \textbf{Scale $s^{\star}$} & \textbf{RAM} &  \textbf{Bootstrap Time} & \textbf{Total Time} & \textbf{$\Omega^{\circ}$}\\
			\midrule
						                                                                        			CCS'18~\cite{tmodel-ccs2018}$^{\dagger}$ & 31\% & 2.6 TiB & 3 days, 11 hrs. & 35 days, 14 hrs. & 1850\\
						                                                                                                                                                			This work$^{\dagger}$ & 31\% & 932 GiB & 17 hrs. & 2 days, 2 hrs. & 79\\
			\midrule
									                                                                                                			This work$^{\ddagger}$ & 100\% & 3.9 TiB & 2 days, 21 hrs. & 8 days, 6 hrs. & 310\\
			\bottomrule
		\end{tabular}
		\begin{tablenotes}
			\item[$\star$] \textbf{31\%}: $\approx$2k relays and $\approx$250k users; \textbf{100\%}: 6,489 relays and 792k users
			\item[$\circ$] $\Omega$: ratio of real time / simulated time in steady state (after bootstrapping)
			\item[$\dagger$] Using 8$\times$10-core Intel Xeon E7-8891v2 CPUs each running @3.2 GHz.
			\item[$\ddagger$] Using 8$\times$18-core Intel Xeon E7-8860v4 CPUs each running @2.2~GHz.
		\end{tablenotes}
	\end{threeparttable}
\vspace{-3mm}
\end{table}

\paragraph{Scalable Simulations}
Our new models and Shadow enhancements enable researchers to run larger
networks faster than was previously possible.
We demonstrate our improvements to scalability in two ways. First, we compare in
the top part of Table~\ref{tab:scalability}
the resources required for the 31\%
experiments described above.
We distinguish total run time from the time required to bootstrap all
Tor relays and clients, initialize all traffic generators, and reach steady
state. We reduced the time required to execute the
bootstrapping process by 2~days, 18~hours, or 80\%, while we reduced the
total time required to run the bootstrapping process plus 25 simulated minutes of steady state by 33 days, 12 hours, or 94\%.
The ratio of real time units required to execute each simulated time unit during steady state
(i.e., after bootstrapping has completed) was reduced by 96\%, further highlighting our achieved speedup.
When compared to models of the same $s=31\%$ scale
from previous work, we observed that our improvements reduced the maximum RAM required
to run bootstrapping plus 25 simulated minutes of steady state from 2.6\,TiB down to 932\,GiB (a total reduction of
1.7\,TiB, or 64\%). 

Second, we demonstrate how our improvements enable us to run significantly
larger models by running three Tor models at scale $s=1.0$, i.e.,
at 100\% of the size of the true Tor network.
We are the first to simulate Tor test networks of this scale.\footnote{We 
	attempted to run a 100\% scale Tor network using the CCS 2018 model~\cite{tmodel-ccs2018},
	but it did not complete the bootstrapping phase within 30 days.}
The bottom part of Table~\ref{tab:scalability} shows
that each of our 100\% Tor networks consumed at most 3.9\,TiB of RAM, completed
bootstrapping in 2 days, 21 hours, and ran the entire simulation (bootstrapping
plus 25 simulated minutes of steady state) in 8 days, 6 hours.
We show in Figure~\ref{fig:torci}
that our 100\% networks also achieve similar
performance compared to the metrics published by Tor~\cite{tormetrics-repro}. Our
results are plotted with 95\% confidence intervals to better understand how
well our sampling methods are capable of reproducing the performance
characteristics of the true Tor network. We describe how to conduct
such a statistical inference in \S\ref{sec:significance} next.

\section{On the Statistical Significance of Results}
\label{sec:significance}

\newcommand\q{\ensuremath{y}}

Recall that our modeling methodology from \S\ref{sec:tormodel} produces
\textit{sampled} Tor networks at scales of $0 < s \leq 1$ times the size of a 100\%
network.
Because these networks are sampled using data from the true Tor network,
there is an associated sampling error that must be quantified when making
predictions about how the effects observed in sampled Tor networks generalize to
the true Tor network.
In this section, we establish a methodology for employing statistical inference
to quantify the sampling error and make useful predictions from sampled
networks.
In our methodology, we: (i)\,use repeated sampling to generate multiple
sampled Tor networks; (ii)\,estimate the true distribution of a random
variable under study through measurements collected from multiple sampled network simulations;
and (iii)\,compute statistical confidence intervals to define
the precision of the estimation.

We remark that it is paramount to conduct a statistical inference when running
experiments in sampled Tor networks in order to contextualize the results they
generate. Our methodology employs confidence intervals (CIs) to establish
the precision of estimations that are made across sampled
networks. CIs will allow a researcher to make a statistical argument about
the extent to which the results they have obtained are relevant to the real world. As we will
demonstrate in \S\ref{sec:casestudy}, CIs help guide researchers to sample
additional Tor networks (and run additional simulations) if necessary for
drawing a particular conclusion in their research.
Our methodology represents a shift in the state of the art
of analysis methods typically used in Tor network performance research, which
has previously ignored statistical inference and CIs altogether (see \S\ref{sec:related:perf}).

\subsection{Methodology}
When conducting research using experimental Tor networks, suppose we have an
interest in a particular network metric; for example, our research might call
for a focus on the distribution of time to last byte across all files of a given
size downloaded through Tor as an indication of Tor performance (see our
ontology in Appendix~\ref{sec:ontol} for examples of other useful metrics).
Because the values of such a variable are determined by the outcomes of
statistical experiments, we refer to the variable as random variable $X$. The
true probability distribution over $X$ is $P(X)$, the true cumulative
distribution is $F_{X}(x) = P(X \leq x)$, and the true inverse distribution at
quantile $\q$ is ${F}_{X}^{-1}(\q)$ such that $\q = F_{X}(F_{X}^{-1}(\q))$. Our
goal is to estimate $P(X)$ (or equivalently, $F_X$ and $F_X^{-1}$), which we do by running
many simulations in sampled Tor networks and averaging the empirical
distributions of $X$ at a number of quantiles across these simulations.
Table~\ref{tab:sigsymbols} summarizes the
symbols that we use to describe our methodology.

\paragraph{Repeated Sampling}
A single \textit{network} sampled from the true Tor network may not consistently produce
perfectly representative results due to the sampling error introduced in the
model sampling process (i.e., \S\ref{sec:netmodel}). Similarly, a single \textit{simulation} may not perfectly
represent a sampled network due to the sampling error introduced by the random choices made in the simulator (e.g., guard selection).
Multiple samples of each are needed to conduct a statistical inference and understand
the error in these sampling processes.

\begin{table}[t]
	\centering
	\footnotesize
	\captionsetup{skip=2pt} 	\begin{threeparttable}
		\caption{Symbols used to describe our statistical methodology.}
		\label{tab:sigsymbols}
		\begin{tabular}{@{\,}c@{\,\,}l@{\,}}
			\toprule
			\textbf{Symbol} & \textbf{Description} \\
			\midrule
			$P(X)$ & true probability distribution of random variable $X$\\
			$F_{X}(x)$ & cumulative distribution function of $X$ at $x$ such that $P(X \leq x)$\\
			${F}_{X}^{-1}(\q)$ & inverse distribution function of $X$ at $\q$ such that $\q = F_{X}(F_{X}^{-1}(\q))$\\
			$\mu(\q)$ & estimate of inverse distribution function at quantile $\q$\\
			$\epsilon(\q)$ & error on inverse distribution estimate at quantile $\q$\\
									\midrule
			$n$ & number of independently sampled Tor networks\\
			$\widehat{P}_{i}(X)$ & probability distribution over $X$ in network $i$\\
			$\widehat{F}_{Xi}(x)$ & cumulative distribution function of $X$ at $x$ such that $\widehat{P_i}(X \leq x)$\\
			$\widehat{F}_{Xi}^{-1}(\q)$ & inverse distribution function of $X$ in network $i$ at quantile $\q$\\
			$\widehat\mu_i(\q)$ & estimate of inverse distribution function in network $i$ at quantile $\q$\\
			$\widehat{\epsilon}_{i}(\q)$ & error on inverse distribution estimate in network $i$ at quantile $\q$\\
						\midrule
			$m_i$ & number of simulations in sampled Tor network $i$\\
			$\nu_{ij}$ & number of samples of $X$ collected from sim $j$ in net $i$\\
			$\widetilde{E}_{ij}(X)$ & empirical distribution over $\nu_{ij}$ samples of $X$ from sim $j$ in net $i$\\
			$\widetilde{F}_{Xij}(x)$ & cumulative distribution function of $X$ at $x$ such that $\widetilde{E_{ij}}(X \leq x)$\\
			$\widetilde{F}_{Xij}^{-1}(\q)$ & inverse distribution function of $X$ from sim $j$ in net $i$ at quantile $\q$\\
			\bottomrule
		\end{tabular}
									\end{threeparttable}
	\vspace{-4mm}
\end{table}

We independently sample $n>0$ Tor networks according to \S\ref{sec:tormodel}.
The $i$th resulting Tor network is associated with a probability distribution
$\widehat{P}_{i}(X)$ which is specific to the $i$th network and the relays that were
chosen when generating it. To estimate $\widehat{P}_{i}(X)$, we run $m_i>0$
simulations in the $i$th Tor network. During the $j$th simulation in the $i$th
network, we sample $\nu_{ij}$~values
of $X$ from $\widehat{P}_{i}(X)$ (i.e., we
collect $\nu_{ij}$ time to last byte measurements from the simulation).
These $\nu_{ij}$ samples form the empirical distribution $\widetilde{E}_{ij}(X)$,
and we have $\sum_{i=1}^{n} m_i$ such distributions in total (one for each simulation).

\paragraph{Estimating Distributions}
Once we have completed the simulations and collected the $\sum_{i=1}^{n} m_i$
empirical distributions, we then estimate the inverse distributions
$\widehat{F}_{Xi}^{-1}$ and $F_{X}^{-1}$ associated with the sampled network and
true probability distributions $\widehat{P}_{i}(X)$ and $P(X)$, respectively.

First, we estimate each $\widehat{F}_{Xi}^{-1}(\q)$ at quantile $\q$ by taking the
mean over the $m_i$ empirical distributions from network $i$:
\begin{equation}\label{eq:estimate1}
	\widehat{F}_{Xi}^{-1}(\q) = \widehat{\mu_i}(\q) = \textstyle\frac{1}{m_i} \textstyle\sum_{j=1}^{m_i} \widetilde{F}_{Xij}^{-1}(\q)
\end{equation}
We refer to $\widehat{\mu_i}$ as an estimator of $\widehat{F}_{Xi}^{-1}$; when
taken over a range of quantiles, it allows us to estimate the cumulative
distribution $\widehat{F}_{Xi}(x) = \widehat{P_i}(X \leq x)$.

Second, we similarly estimate $F_{X}^{-1}$ over all networks by taking the mean
over the $n$ distributions estimated above:
\begin{equation}\label{eq:estimate2}
	F_{X}^{-1}(\q) \approx \mu(\q) = \textstyle\frac{1}{n} \textstyle\sum_{i=1}^{n} \widehat{\mu}_i(\q)
\end{equation}
We refer to $\mu$ as an estimator of $F_{X}^{-1}$; when
taken over a range of quantiles, it allows us to estimate the cumulative
distribution $F_{X}(x) = P(X \leq x)$.

\begin{figure}[t]
	\captionsetup{skip=0pt} 	\begin{subfigure}[b]{0.49\columnwidth}
		\centering
		\captionsetup{skip=0pt} 		\includegraphics[width=1.0\textwidth]{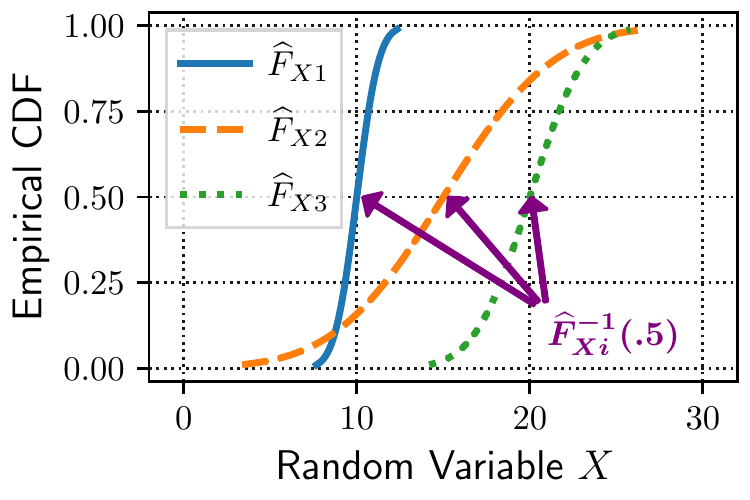}
		\caption{}
		\label{fig:sigexampleecdf}
	\end{subfigure}
	\begin{subfigure}[b]{0.49\columnwidth}
		\centering
		\captionsetup{skip=0pt} 		\includegraphics[width=1.0\textwidth]{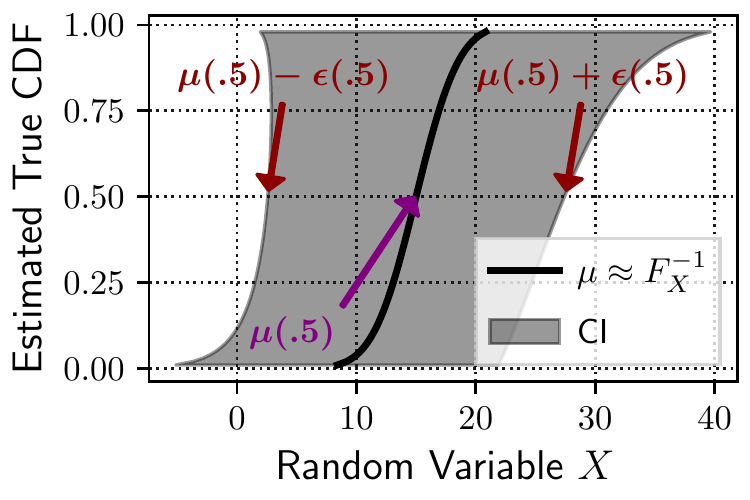}
		\caption{}
		\label{fig:sigexampleci}
	\end{subfigure}
	\caption{A synthetic example of estimating the cumulative distribution of a
		random variable $X$ (e.g., time to last byte).
		\subref{fig:sigexampleecdf} The mean in Equation~\ref{eq:estimate2} and standard deviation in Equation~\ref{eq:error2} are computed over the $n=3$ values at each quantile.
		\subref{fig:sigexampleci} The estimated true distribution from Equation~\ref{eq:estimate2} is shown with confidence intervals from Equation~\ref{eq:ci}.
	}
	\label{fig:sigexample}
	\vspace{-3mm}
\end{figure}

We visualize the process of estimating $F_{X}^{-1}$ in
Figure~\ref{fig:sigexample} using an example: Figure~\ref{fig:sigexampleecdf}
shows $n=3$ synthetic distributions where the upward arrows point to
the $\widehat{F}_{Xi}^{-1}$ values from network $i$ at quantile $\q=.5$, and
Figure~\ref{fig:sigexampleci} shows the mean of those values as the estimator
$\mu$. The example applies analogously when estimating each $\widehat{F}_{Xi}^{-1}$.

\paragraph{Computing Confidence Intervals}
We quantify the precision of our estimator $\mu$ using CIs.
To compute the CIs, we first quantify the measurement error associated with the empirical samples.  This will often be negligible, but a possible source of nontrivial measurement error is resolution error; that is, if the empirical results are reported to a resolution of $r$ (e.g., 0.01\,s), the resolution error for each sample will be $\frac{r}{\sqrt{12}}$, and the resolution error ${\zeta_i}$ for the empirical mean $\widehat{\mu_i}(\q)$ of network $i$ at quantile $\q$ is ${\zeta_i} = \frac{r}{\sqrt{12 m_i}}$.
Next, we quantify the sampling error associated with the
estimates from Equations~\ref{eq:estimate1} and~\ref{eq:estimate2}. The error
associated with $\widehat{\mu}_i$ for network $i$ at quantile $\q$ is:
\begin{equation}\label{eq:error1}
	\widehat{\epsilon}_{i}(\q) = \widehat{\sigma}_i(\q) \cdot t / \sqrt{m_i-1}
\end{equation}
where $\widehat{\sigma}_i(\q) = \sqrt{\frac{1}{m_i} \sum_{j=1}^{m_i} ( \widetilde{F}_{Xij}^{-1}(\q) - \widehat{\mu_i}(\q) )^2 + \zeta_i^2}$
is the standard deviation over the $m_i$ empirical values at quantile $\q$
(including the measurement error)
and $t$ is
the $t$-value from the Student's $t$-distribution at confidence level $\alpha$
with $m_i-1$ degrees of freedom~\cite[\S10.5.1]{statsbook}. $\widehat{\epsilon}_{i}(\q)$
accounts for the sampling error and estimated true variance of the underlying
distribution at~$\q$.
The error associated with $\mu$ at quantile $\q$ is:
\begin{equation}\label{eq:error2}
	\epsilon(\q) = \delta(\q) + \sigma(\q) \cdot t / \sqrt{n-1}
\end{equation}
where
$\sigma(\q) = \sqrt{\frac{1}{n} \sum_{i=1}^{n} ( \widehat{F}_{Xi}^{-1}(\q) - \mu(\q) )^2}$
is the standard deviation over the $n$ estimated inverse distribution values at quantile $\q$, and
$ \delta(\q) = \frac{1}{n} \sum_{i=1}^{n} \widehat{\epsilon}_{i}(\q)$
is the mean error from $\widehat{\mu}_i$ over all $n$ sampled networks.
$\epsilon(\q)$ accounts for the sampling error introduced in the Tor network model generation and in the simulations.
We can then define the CI at quantile $\q$ as the interval that contains the
true value from the inverse distribution $F_{X}^{-1}(\q)$ with probability
$\alpha$:
\begin{equation}\label{eq:ci}
\mu(\q) - \epsilon(\q) \leq F_{X}^{-1}(\q) \leq \mu(\q) + \epsilon(\q)
\end{equation}
The width of the interval is $2\cdot\epsilon(\q)$, which we visualize at $\q=.5$
with the downward arrows and over all quantiles with the shaded region in
Figure~\ref{fig:sigexampleci}.

\subsection{Discussion}

\paragraph{Number of Samples Per Simulation}
Recall that we collect $\nu_{ij}$
empirical samples of the random variable $X$
from simulation $j$ in network $i$.
If we increase $\nu_{ij}$ (e.g., by running
the simulation for a longer period of time), this will result in a ``tighter''
empirical distribution $\widetilde{E}_{ij}(X)$ that will more closely resemble the probability
distribution $\widehat{P}_i(X)$. However, from Equation~\ref{eq:estimate1} we can see
that $\widetilde{E}_{ij}(X)$ only contributes a single value to the computation of
$\widehat{\mu}_i$ for each quantile. Therefore, once we have enough samples
so that $\widetilde{E}_{ij}(X)$ reasonably approximates $\widehat{P}_i(X)$,
it is more useful to run new simulations than to gather additional
samples from the same simulation.

\looseness-1
\paragraph{Number of Simulations Per Network}
Additional simulations in network $i$ will provide us with additional empirical
distributions $\widetilde{E}_{i*}(X)$, which will enable us to obtain a better
estimate of $\widehat{P}_i(X)$. Moreover, it will also increase the precision of the CI
by reducing $\widehat{\epsilon}_i$ in Equation~\ref{eq:error1}: increasing the
number of $\widetilde{E}_{i*}(X)$ values at each quantile will decrease the standard
deviation $\widehat{\sigma}_i$ (if the values are normally distributed) and the
$t$-value (by increasing the number of degrees of freedom) while increasing the square root
component (in the denominator of $\widehat{\epsilon}_i$).

\begin{figure}[t]
	\captionsetup{skip=0pt} 	\centering
	\includegraphics[width=\singlefigwidthfactor\columnwidth]{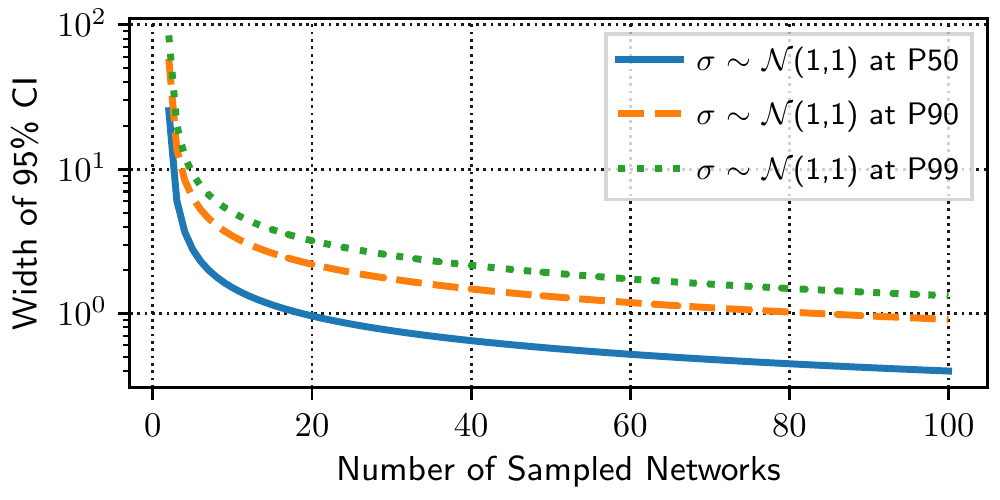}
	\caption{
The width of the 95\% CI (on the log-scale y-axis) can be significantly reduced
by more than an order of magnitude after running experiments in fewer than 10
independently sampled Tor networks (when $\sigma$ is normally distributed
according to $\mathcal{N}(1,1)$).
	}
	\label{fig:ciwidth}
	\vspace{-3mm}
\end{figure}

\paragraph{Number of Sampled Networks}
Additional simulations in independently sampled Tor networks will provide us
with additional estimated $\widehat{P}_i(X)$ distributions, which will enable us to
obtain a better estimate of $P(X)$. Similarly as above, additional $\widehat{P}_i(X)$
estimates will increase CI precision by reducing $\epsilon$
in Equation~\ref{eq:error2}: 
the standard deviation $\sigma$ and the $t$-value will decrease
while the square root component will increase.

To give a concrete example, suppose
$\sigma$ is normally distributed
according to $\mathcal{N}(1,1)$. The
width of the resulting CI for each number of sampled networks $n \in [2,100]$ at
quantiles $\q \in \{0.5,0.9, 0.99\}$ (i.e., P50, P90, and P99, respectively) is
shown in Figure~\ref{fig:ciwidth}. Notice that the y-axis is drawn at log-scale,
and shows that the width of the CI can be significantly reduced by more than an
order of magnitude after running experiments in even just a small number of
sampled networks. Additionally, we can see that the main improvement in
confidence results from the first ten or so sampled networks, after which we
observe relatively diminishing returns.

\paragraph{Scale}
Another important factor to consider is the network scale $0 < s \leq 1$. Larger
scales $s$ (closer to 1) cause the probability distribution $\widehat{P}_i(X)$ of
each sampled network to cluster more closely around the true probability
distribution $P(X)$, while smaller values cause the $\widehat{P}_i(X)$ to vary more
widely. Larger scales $s$ therefore induce smaller values of $\sigma(\q)$
and therefore $\epsilon(\q)$. (See \S\ref{sec:results} for a demonstration of
this phenomenon.)

\begin{figure}[t]
	\captionsetup{skip=0pt} 	\centering
	\includegraphics[width=\singlefigwidthfactor\columnwidth]{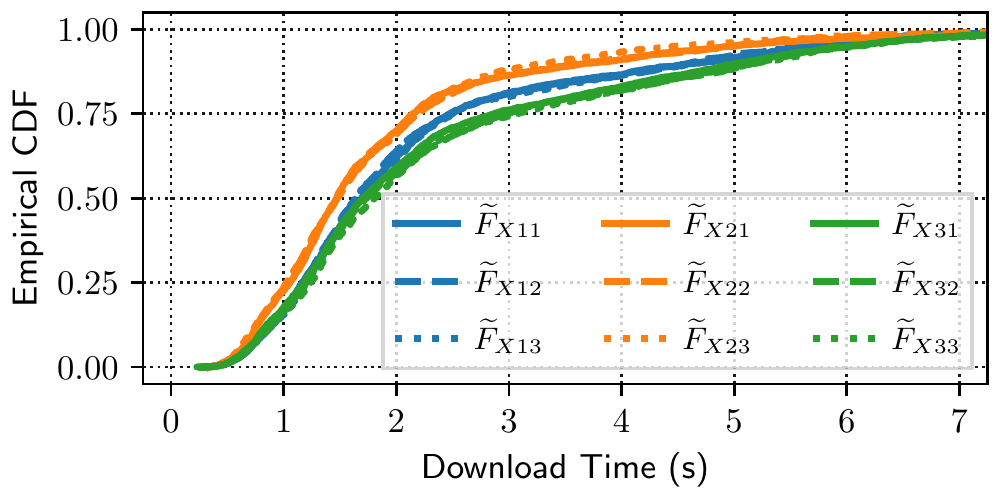}
	\caption{
		Sampling error introduced by Shadow is much less significant than error
		introduced by Tor network sampling (i.e., \S\ref{sec:netmodel}).
	}
	\label{fig:seedsvsnets}
	\vspace{-3mm}
\end{figure}

\paragraph{Sampling Error in Shadow}
While $\epsilon$ includes the error due to sampling a scaled-down Tor network
(i.e., \S\ref{sec:netmodel}), the main error that is accounted for in
$\widehat{\epsilon}_i$ is the sampling error introduced by the choices made in the
simulator. If this error is low, running additional simulations in the same
network will have a reduced effect. To check the sampling error introduced by
Shadow, we ran 9 simulations (3 simulations in each of 3 independently sampled
networks of scale $s=0.1$) with unique simulator seeds.
Figure~\ref{fig:seedsvsnets} shows that the empirical distributions of the
50~KiB download times vary much more widely across sampled Tor networks
than they do across simulations in the same network.
Although it is ideal to run multiple simulations in each of multiple sampled
networks, our results indicate that it may be a better use of resources to run
\textit{every} simulation in an independently sampled network. We believe this to be a
reasonable optimization if a lack of available computational resources is a concern.

\paragraph{Conclusions}
We have established a methodology for estimating the true distribution of
random variables being studied across simulations in multiple Tor networks.
Importantly, our methodology includes the computation of CIs that help
researchers make statistical arguments about the conclusions they draw from Tor
experiments.
As we explained above and demonstrated in Figure~\ref{fig:ciwidth}, running
simulations in smaller-scale Tor networks or in a smaller number of Tor networks
for a particular configuration leads to larger CIs that limit us to drawing
weaker conclusions from the results. Unfortunately, previous Tor research that
utilizes Tor networks has focused exclusively on
single Tor networks while completely ignoring CIs, leading to questionable
conclusions (see \S\ref{sec:related:perf}).We argue that our methodology is superior to the state-of-the-art
methods, and present in~\S\ref{sec:casestudy} a case study demonstrating how to
put our methods into practice while conducting Tor research.

\section{Case Study: Tor Usage and Performance}
\label{sec:casestudy}

This section presents a case study on the effects of an increase in
Tor usage on Tor client performance. Our
primary goal is to demonstrate how to apply the methodologies we presented
throughout this paper through a concrete set of experiments.

\subsection{Motivation and Overview}
\looseness-1
Growing the Tor network is desirable because it improves
anonymity~\cite{econymics} and access to information online~\cite{freeinfo}.
One strategy for facilitating wider adoption of
Tor is to deploy it in more commonly used browsers. Brave
now prominently advertises on its website Tor integration
into its browser's private browsing mode, giving
users the option to open a privacy-enhanced tab that routes
traffic through Tor~\cite{bravesite}, and Mozilla is also interested
in providing a similar ``Super Private Browsing'' mode for Firefox
users~\cite{mozillacallcfp}. However, Tor has never been deployed at
the scale of popular browser deployments (Firefox has >250M monthly active users~\cite{mozillascale}), 
and many important research problems must be considered
before such a deployment could occur~\cite{mozillacallblog}.
For example, deploying Tor more widely could add enough load to the network
that it reduces performance to the extent that some users are dissuaded from using it~\cite{privatelywaiting}
while reducing anonymity for those that remain~\cite{econymics}.

There has been little work in understanding the performance effects of
increasing Tor network load as representative of the
significant change in Tor usage that would likely occur in a wider deployment. Previous work that considered variable load did so
primarily to showcase a new simulation
tool~\cite{jansen2012shadow} or to inform the design of a particular
performance-enhancing
algorithm~\cite{kist-tops2018,jansen2012throttling}
rather than for the purpose of understanding network growth and
scalability~\cite{walkingonions}. Moreover, previous studies of the effects of load on performance
lack analyses of the statistical significance of the reported results, raising
questions as to their practical meaning.

\looseness-1
Guided by the foundations that we set out in this paper, we explore the
performance effects of a sudden rise in Tor usage that could result from, e.g.,
a Mozilla deployment of Tor. In particular, we demonstrate the use of our
methodologies with an example study of this simple hypothesis: 
\textit{increasing the total user traffic load in Tor by 20\% will
reduce the performance of existing clients by increasing their download
times and download error rates}. To study this hypothesis,
we conduct a total of 420 simulations in independently sampled Tor networks across three
network scale factors and two traffic load factors; we measure relevant
performance properties and
conduct a statistical analysis of the results following our methodology
in~\S\ref{sec:significance}. Our study demonstrates how to use our
contributions to conduct statistically valid Tor performance research.

\subsection{Experiment Setup}
\label{sec:load:setup} 
\paragraph{Experiments and Simulations}
We refer to an \textit{experiment} as a unique pair of network scale $s$ and load $\load$
configurations, and a \textit{simulation}
as a particular execution of an experiment configuration. We study our hypothesis
with a set of 6 experiments; for each experiment, we run multiple simulations
in independent Tor networks
so that we can quantify the statistical significance of the results following our
guidance from~\S\ref{sec:significance}.

\paragraph{Tor Network Scale and Load}
The Tor network scales that a researcher can consider are typically dependent on
the amount of RAM to which they have access. Although we were able to run a
100\% Tor network for our evaluation in \S\ref{sec:platforms}, we do not expect
that access to a machine with 4~TiB of RAM, as was required to
run the simulation, will be common. Because it will be more informative, we focus our study
on multiple smaller network scales with more accessible resource requirements
while showing the change in confidence that results from running networks of
different scales.
In particular, our study considers Tor network scales of 1\%,
10\%, and 30\% ($s \in \{0.01,0.1,0.3\}$) of the size of the true Tor network.
At each of these network scales, we study the
performance effects of 100\% and 120\% traffic load ($\load \in \{1.0,1.2\}$)
using a process scale factor of $p=0.01$, i.e., each TGen process simulates
$1/0.01=100$ Tor users.

\begin{table}[t]
	\centering
	\captionsetup{skip=2pt} 	\footnotesize
	\begin{threeparttable}
		\caption{Tor usage and performance experiments in Shadow}
		\label{tab:loaddetails}
		\begin{tabular}{c@{\,\,\,\,\,\,}c@{\,\,\,\,\,\,}c@{\,\,\,\,\,\,}c|c@{\,\,\,\,\,\,}c}
			\toprule
			\textbf{Scale $s$} & \textbf{Load $\load$} & \textbf{Sims $n$} & \textbf{CPU$^{\star}$} & \textbf{RAM/Sim$^{\dagger}$} & \textbf{Run Time/Sim$^{\ddagger}$} \\
			\midrule
			1\% & 100\% & 100 & 4$\times$8 & 35 GiB & 4.8 hours\\
			1\% & 120\% & 100 & 4$\times$8 & 50 GiB & 6.7 hours\\
			\midrule
			10\% & 100\% & 100 & 4$\times$8 & 355 GiB & 19.4 hours\\
			10\% & 120\% & 100 & 4$\times$8 & 416 GiB & 23.4 hours\\
			\midrule
			30\% & 100\% & 10 & 8$\times$8 & 1.07 TiB & 4 days, 21 hours\\
			30\% & 120\% & 10 & 8$\times$8 & 1.25 TiB & 5 days, 22 hours\\
			\bottomrule
		\end{tabular}
		\begin{tablenotes}
						\item[$\star$] \textbf{4$\times$8}-core Intel Xeon E5 @3.3 GHz; \textbf{8$\times$8}-core Intel Xeon E5 @2.7 GHz.
			\item[$\dagger$] The median of the per-simulation max RAM usage over all simulations.
			\item[$\ddagger$] The median of the per-simulation run time over all simulations.
		\end{tablenotes}
	\end{threeparttable}
	\vspace{-2mm}
\end{table}

\paragraph{Number of Simulations}
Another important consideration in our evaluation is the number $n$ of
simulations to run for each experiment. As explained
in~\S\ref{sec:significance}, running too few simulations will result in wider
confidence intervals that will limit us to weaker conclusions. The number $n$ of
simulations that should be run typically depends on the results and the
arguments being made, but in our case we run more than we require to validate our
hypothesis in order to demonstrate the effects of varying $n$.
As shown in the left part of Table~\ref{tab:loaddetails}, we run a total of 420
simulations across our 6 experiments (three network scales and two load factors) using
two machine profiles: one profile included 4$\times$8-core Intel Xeon E5-4627
CPUs running at a max clock speed of 3.3\,GHz and 1.25\,TiB of RAM; the other
included 8$\times$8-core Intel Xeon E5-4650 CPUs running at a max clock speed of
2.7\,GHz and 1.5\,TiB of RAM.

\begin{table}[t]
	\centering
	\captionsetup{skip=2pt} 	\footnotesize
	\begin{threeparttable}
		\caption{Network composition in each simulation$^{\star}$}
		\label{tab:netdetails}
		\begin{tabular}{@{\,\,}c@{\,\,}|@{\,\,}c@{\,\,}c@{\,\,}c@{\,\,}c@{\,\,}c@{\,\,}|@{\,\,}c@{\,\,}c@{\,\,}|@{\,\,}c@{\,\,}}
			\toprule
			\textbf{Scale $s$} & \textbf{DirAuth} & \textbf{Guard} & \textbf{Middle} & \textbf{Exit} & \textbf{E+G}$^{\dagger}$ & \textbf{Markov} & \textbf{Perf}$^{\ddagger}$ & \textbf{Server}\\
			\midrule
			1\% & 3 & 20 & 36 & 4 & 4 & 100 & 8 & 10 \\
			10\% & 3 & 204 & 361 & 40 & 44 & 792 & 79 & 79 \\
			30\% & 3 & 612 & 1,086 & 118 & 129 & 2,376 & 238 & 238 \\
			\bottomrule
		\end{tabular}
		\begin{tablenotes}
			\item[$\star$] \textbf{Total} number of relays at $s$=1\%: 67; at $s$=10\%: 652; and at $s$=30\%: 1,948.
			\item[$\dagger$] \textbf{E+G}: Relays with both the exit and guard flags $^{\ddagger}$ \textbf{Perf}: Benchmark clients
		\end{tablenotes}
	\end{threeparttable}
\end{table}

\begin{figure*}[t]
	\centering
	\captionsetup{skip=2pt} 	\begin{subfigure}[b]{0.33\textwidth}
		\centering
		\captionsetup{skip=0pt} 		\includegraphics[width=1.0\textwidth]{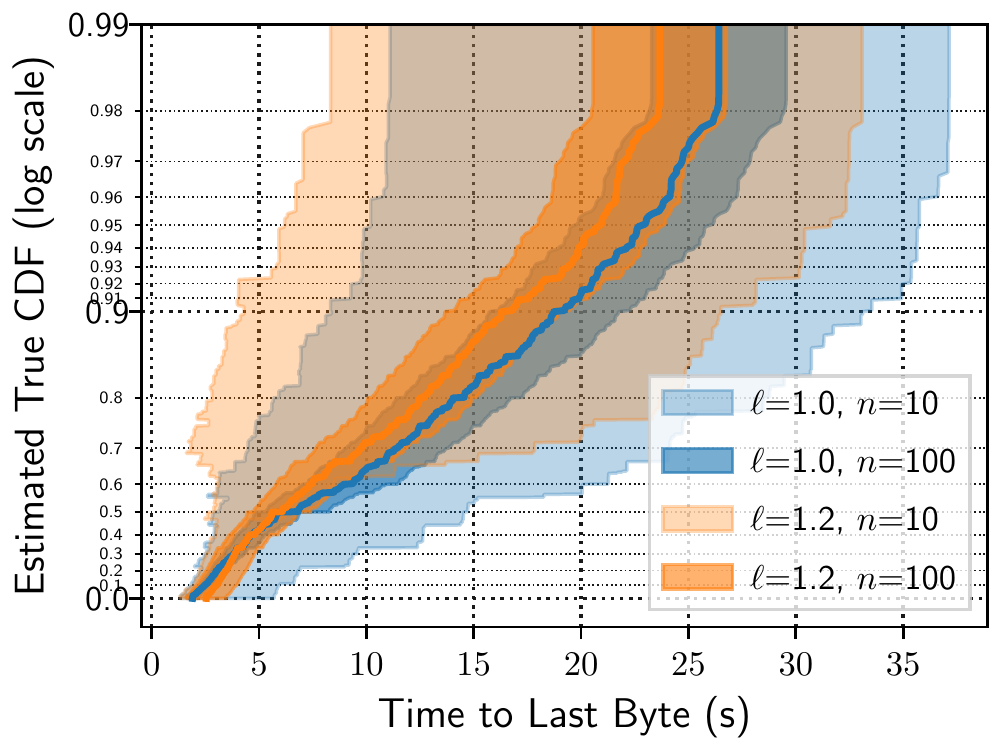}
		\caption{1\% Network Scale ($s = 0.01$)}
		\label{fig:ttlb:1m:one}
	\end{subfigure}
	\begin{subfigure}[b]{0.33\textwidth}
		\centering
		\captionsetup{skip=0pt} 		\includegraphics[width=1.0\textwidth]{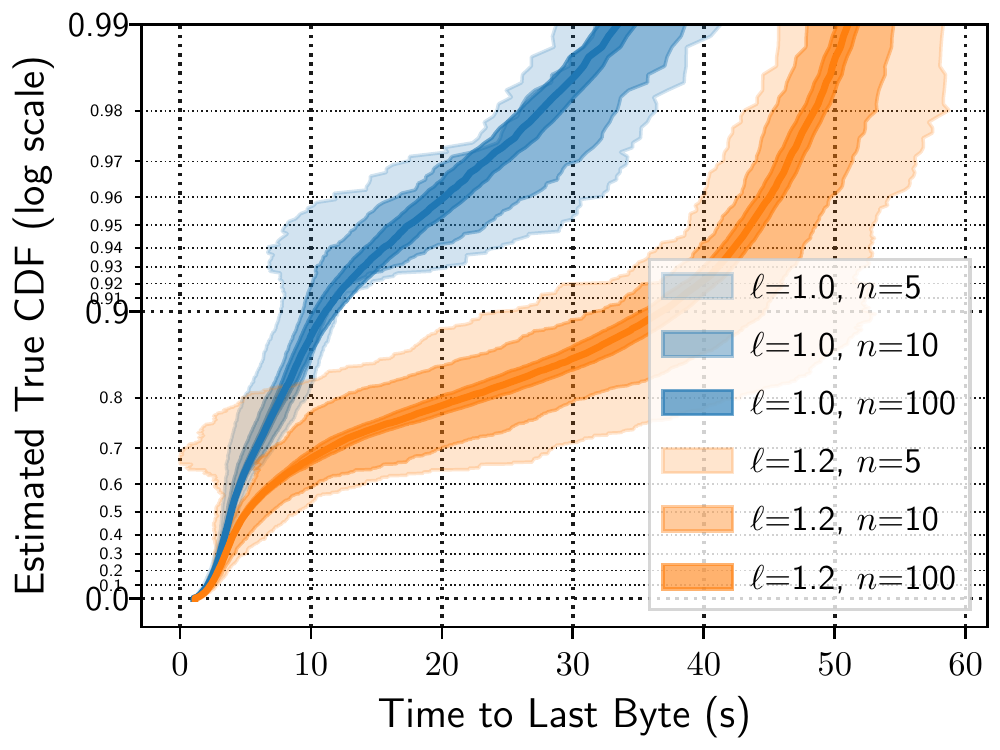}
		\caption{10\% Network Scale ($s = 0.1$)}
		\label{fig:ttlb:1m:ten}
	\end{subfigure}
	\begin{subfigure}[b]{0.33\textwidth}
		\centering
		\captionsetup{skip=0pt} 		\includegraphics[width=1.0\textwidth]{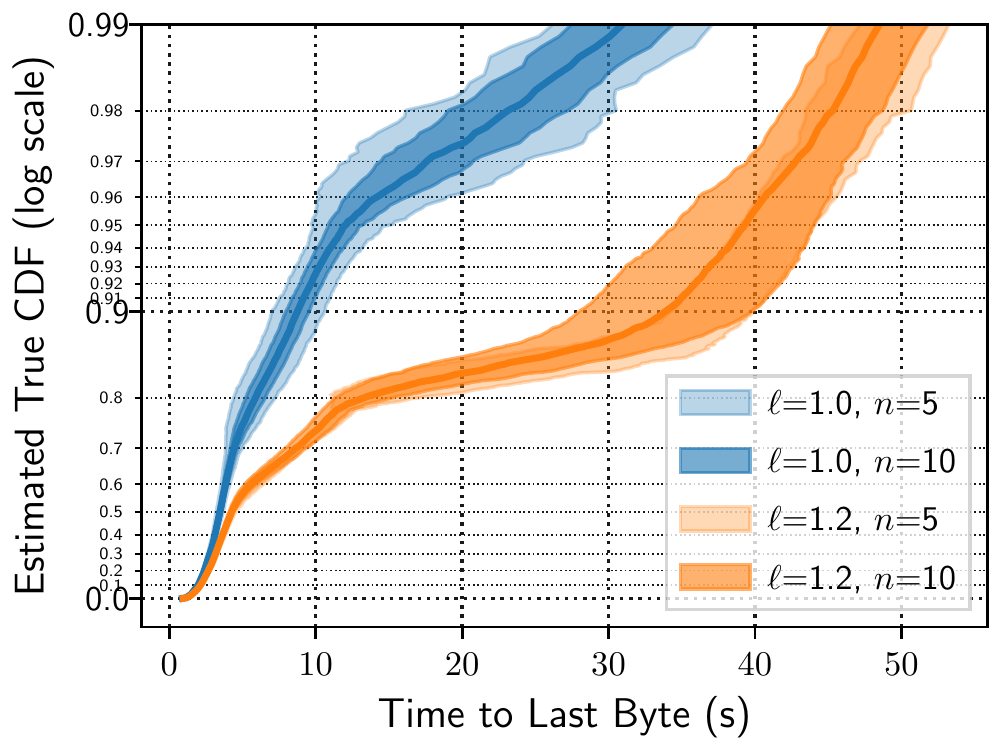}
		\caption{30\% Network Scale ($s = 0.3$)}
		\label{fig:ttlb:1m:thirty}
	\end{subfigure}
	\caption{
		Time to last byte in seconds of 1~MiB downloads from performance benchmarking
		clients from experiments with traffic load $\load=1.0$ and $\load=1.2$ in
		networks of various scale $s$. The results from each experiment are aggregated
		from $n$ simulations following \S\ref{sec:significance}, and the CDFs are
		plotted with tail-logarithmic y-axes in order to highlight the long tail of network
		performance.
	}
	\label{fig:ttlb:1m}
	\vspace{-3mm}
\end{figure*}

\paragraph{Simulation Configuration}
We run each simulation using an independently sampled Tor network in order to ensure that we
produce informative samples following our guidance from~\S\ref{sec:significance}.
Each Tor network is generated following our methodology from
\S\ref{sec:netmodel} using the parameter values described above
and Tor network state files from January 2019.
The resulting network composition for each scale $s$ is shown in Table~\ref{tab:netdetails}.

Each simulation was configured to run for 1 simulated hour. The relays
bootstrapped a Tor overlay network within the first 5 minutes; all of
the TGen clients and servers started their traffic generation process
within 10 simulated minutes of the start of each simulation. TGen
streams created by Markov clients were set to time out if no bytes were transferred in any
contiguous 5 simulated minute period (the default apache client
timeout), or if the streams were not complete within an absolute time of
10 simulated minutes. Timeouts for streams created by benchmarking clients were set to
15, 60, and 120 seconds for 50\,KiB, 1\,MiB, and 5\,MiB transfers,
respectively.

\subsection{Results}
\label{sec:results}

During each simulation, we measure and collect the properties
that allow us to understand our hypothesis. Ultimately, we
would like to test if increasing the traffic load on the network by 20\% (from
$\load=1.0$ to $\load=1.2$) will reduce client performance. Therefore, we focus
this study on client \textit{download time} and \textit{download error rates} while noting that it
will very likely be useful to consider additional properties when studying more complex
hypotheses (see Appendix~\ref{sec:ontol}).

For each experiment, we combine the results from the $n$ simulations\footnote{We ignore
	the results from the first 20~simulated minutes of each simulation to allow time
	for the network to bootstrap and reach a steady state.} following
the methodology outlined in~\S\ref{sec:significance} and present the estimated
true cumulative distributions with the associated CIs
(as in Figure~\ref{fig:sigexample}) at $\alpha=95\%$ confidence.
We plot the results for varying values of $n$ as
overlapping intervals (the CIs tighten as $n$ increases) for instructional
purposes. Finally, we compare our results across network scales $s$ to highlight
the effect of scale on the confidence in the results.

\begin{figure*}[t]
	\centering
	\captionsetup{skip=2pt} 	\begin{subfigure}[b]{0.33\textwidth}
		\centering
		\captionsetup{skip=0pt} 		\includegraphics[width=1.0\textwidth]{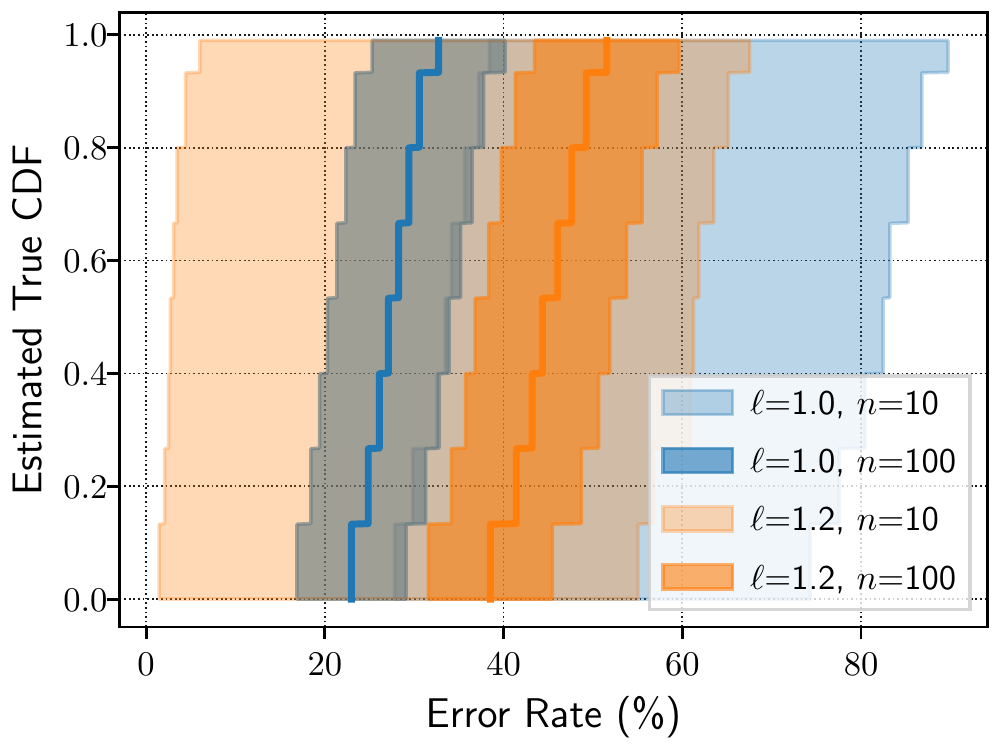}
		\caption{1\% Network Scale ($s = 0.01$)}
		\label{fig:tgenerr:one}
	\end{subfigure}
	\begin{subfigure}[b]{0.33\textwidth}
		\centering
		\captionsetup{skip=0pt} 		\includegraphics[width=1.0\textwidth]{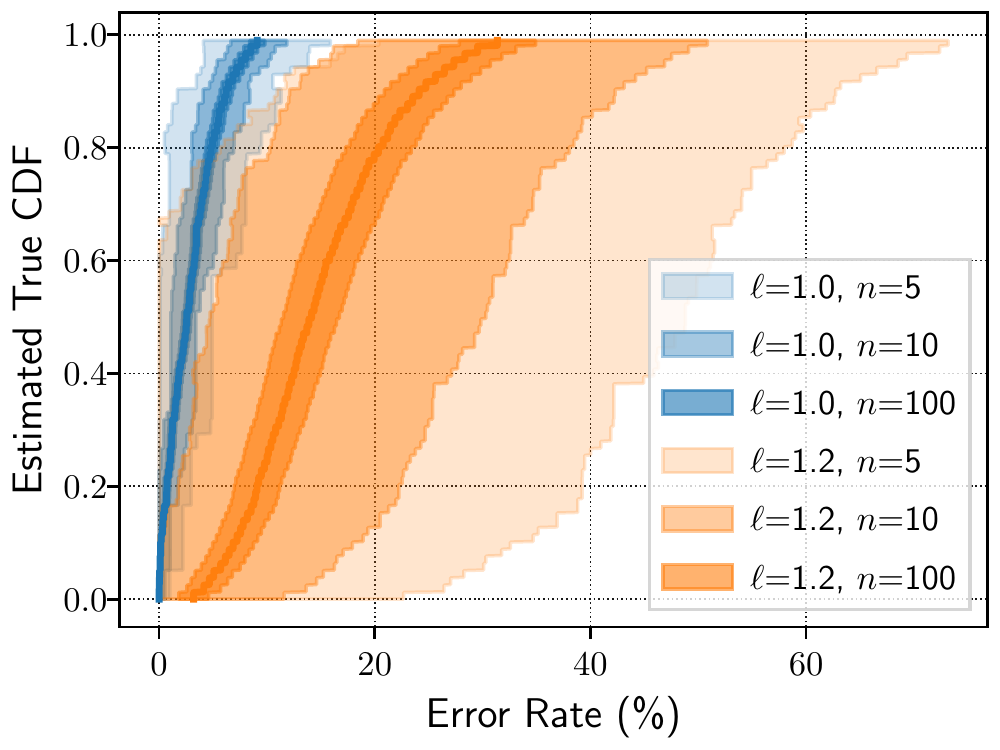}
		\caption{10\% Network Scale ($s = 0.1$)}
		\label{fig:tgenerr:ten}
	\end{subfigure}
	\begin{subfigure}[b]{0.33\textwidth}
		\centering
		\captionsetup{skip=0pt} 		\includegraphics[width=1.0\textwidth]{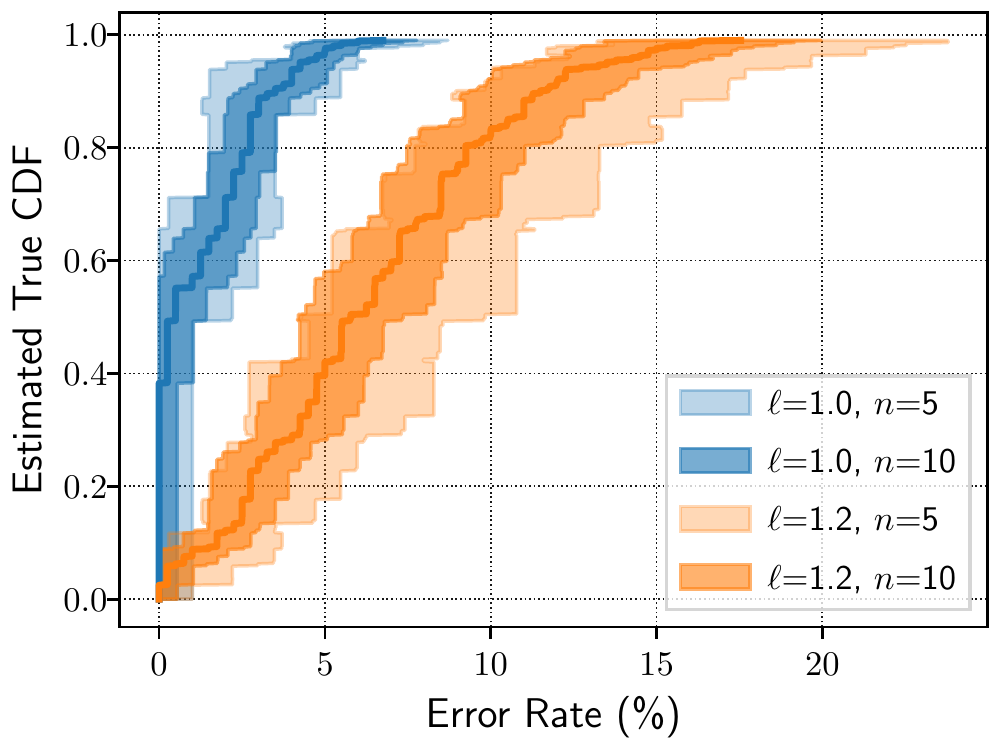}
		\caption{30\% Network Scale ($s = 0.3$)}
		\label{fig:tgenerr:thirty}
	\end{subfigure}
	\caption{
		The download error rate (i.e., the fraction of failed over attempted
		downloads) for downloads of all sizes from performance benchmarking
		clients from experiments with traffic load $\load=1.0$ and $\load=1.2$ in
		networks of various scale $s$. The results from each experiment are aggregated
		from $n$ simulations following \S\ref{sec:significance}.
	}
	\label{fig:tgenerr}
\end{figure*}

\paragraph{Client Download Time}
The time it takes to download a certain number of bytes through Tor (i.e., the
time to first/last byte) allows us to assess and compare
the overall performance that a Tor client experiences. We measure download times
for the performance benchmarking clients throughout the simulations. We present
in Figure~\ref{fig:ttlb:1m} the time to last byte for 1~MiB file downloads, while
noting that we find similar trends for other file download sizes as shown in
  Figures~\ref{fig:ttfb},~\ref{fig:ttlb:50k}, and~\ref{fig:ttlb:5m} in
  Appendix~\ref{app:perf}.
The CDFs are plotted with tail-logarithmic y-axes
in order
to highlight the long tail of network performance as is typically used as an
indication of usability.

Figure~\ref{fig:ttlb:1m:one}
shows the result of our statistical analysis
from~\S\ref{sec:significance} when using a network scale of 1\% ($s=0.01$).
Against our expectation, our estimates of the true CDFs (i.e., the solid lines)
indicate that the time to download 1~MiB files actually \textit{decreased} after
we increased the traffic load by 20\%. However, notice the extent to which the
confidence intervals overlap: for example, the width of the region of overlap of
the $\load=1.0$ and $\load=1.2$ CIs is about 20 seconds at P90 (i.e., at $x \in
[8, 28]$ seconds) when $n=10$, and is about 3 seconds at P90 (i.e., at $x \in
[16.5, 19.5]$ seconds) when $n=100$. Importantly, the estimated true CDF for
$\load=1.0$ falls completely within the CIs for $\load=1.2$ and the estimated
true CDF for $\load=1.2$ falls completely within the CIs for $\load=1.0$, even
when considering $n=100$ simulations for each experiment. Therefore, it is
possible that the $x$ position of the true CDFs could actually be swapped
compared to what is shown in Figure~\ref{fig:ttlb:1m:one}. If we had followed
previous work and ignored the CIs, it would have been very difficult to notice
this statistical possibility. Based on these results alone, we are unable to draw
conclusions about our hypothesis at the desired confidence.

Our experiments with the network scale of 10\% offer more reliable results.
Figure~\ref{fig:ttlb:1m:ten} shows the extent to which the CIs become narrower
as $n$ increases from 5 to 10 to 100. Although there is some overlap in the
$\load=1.0$ and $\load=1.2$ CIs at some $y < 0.9$ values when $n$ is either 5 or
10, we can confidently confirm our hypothesis when $n=100$ because the estimated
true CDFs and their CIs are completely distinguishable. Notice that the CI
precision at $n=10$ and $n=100$ has increased compared to those from
Figure~\ref{fig:ttlb:1m:one}, because the larger scale network
produces more representative empirical samples.

Finally, the results from our experiments with the network scale of 30\%
reinforce our previous conclusions about our hypothesis.
Figure~\ref{fig:ttlb:1m:thirty} shows that the estimated true CDFs and their CIs
are completely distinguishable, allowing us to confirm our hypothesis even when
$n=5$. However, we notice an interesting phenomenon with the $\load=1.2$ CIs:
the CI for $n=10$ is unexpectedly \textit{wider} than the CI for $n=5$. This can
be explained by the analysis shown in Figure~\ref{fig:ciwidth}: as $n$
approaches 1, the uncertainty in the width of the CI grows rapidly.
In our
case, the empirical distributions from the first $n=5$ networks that we
generated happened to be more closely clustered by chance, but $n=10$
resulted in a more diverse set of sampled networks that produced more varied
empirical distributions. Our conclusions happen to be the same both when $n=5$
and when $n=10$, but this may not always be the case (e.g., when the performance
differences between two experiments are less pronounced). We offer general
conclusions based on our results later in this section.

\paragraph{Client Download Error Rate}
The client download error rate (i.e., the fraction of failed over attempted
downloads) helps us understand how additional traffic load would impact
usability. Larger error rates indicate a more congested network and represent a
poorer user experience. We measure the number of attempted and failed downloads
throughout the simulations, and compute the download error rate across all
downloads (independent of file size) for each performance benchmarking client.
We present in Figure~\ref{fig:tgenerr} the download error rate across all
benchmarking clients. (Note that another general network assessment
metric, Tor network goodput, is shown in
  Figure~\ref{fig:oniontracegput} in Appendix~\ref{app:perf}.)

Figure~\ref{fig:tgenerr:one} shows the result of our statistical analysis
from~\S\ref{sec:significance} when using a network scale of 1\% ($s=0.01$). As
with the client download time metric, we see overlap in the $\load=1.0$ and
$\load=1.2$ CIs when $n=10$.
Although it appears that the
download error rates \textit{decrease} when adding 20\% load (because the range
of the $\load=1.0$ CI is generally to the right of the range of the $\load=1.2$ CI),
we are unable to draw conclusions at the desired confidence when $n=10$.
However, the $\load=1.0$ and $\load=1.2$ CIs become significantly narrower
(and no longer overlap) with $n=100$ simulations,
and it becomes clear that adding 20\% load increases the error rate.

Our experiments with the network scale of 10\% again offer more reliable
results. Figure~\ref{fig:tgenerr:ten} shows significant overlap in the
$\load=1.0$ and $\load=1.2$ CIs when $n=5$ simulations and a very slight overlap
in CIs when $n=10$. However, based on the estimated true CDF and CIs when
$n=100$, we can again confidently conclude that increasing the
traffic load by 20\% increases the download error rate because the CIs are
clearly distinguishable. Notice that the CI precision for $\load=1.2$
compared to the CI precision for $\load=1.0$ offers an additional insight into
the results: the error rate is more highly varied when $\load=1.2$, indicating
that the user experience is much less \textit{consistent} than it is when
$\load=1.0$.

Finally, the results from our experiments with the network scale of 30\% again
reinforce our previous conclusions about our hypothesis.
Figure~\ref{fig:tgenerr:thirty} shows that the estimated true CDFs and their CIs
are completely distinguishable, allowing us to confirm our hypothesis even when
$n=5$.

\paragraph{Conclusions}
We offer some general observations based on the results of our case study.
First, our results indicate that it is possible to come to similar conclusions
by running experiments in networks of different scales. Generally, fewer
simulations will be required to achieve a particular CI precision in networks of
larger scale than in networks of smaller scale.
The network scale that is appropriate and the precision that is needed will
vary and depend heavily on the experiments and metrics being compared and the
hypothesis being tested.
However, based on our results, we suggest that networks at a scale of at least
10\% ($s \geq 0.1$) are used whenever possible, and we strongly recommend that
1\% networks be avoided due to the unreliability of the results they generate.
Second, some of our results exhibited the phenomenon that increasing the number of
simulations $n$ also decreased the CI precision, although the opposite is expected.
This behavior is due to random sampling and is more likely
to be exhibited for smaller $n$.
Along with the analysis from \S\ref{sec:significance}, our results lead us to
recommend that no fewer than $n=10$ simulations be run for any
experiment, independent of the network scale $s$.

\section{Conclusion} \label{sec:conc}

In this paper, we develop foundations upon which future Tor
performance research can build.
The foundations we develop include:
(i)\,a new Tor network modeling methodology and supporting tools that produce
\textit{more representative} Tor networks~(\S\ref{sec:netmodel});
(ii)\,accuracy and performance improvements to the Shadow simulator that allow us
to run Tor simulations \textit{faster} and at a \textit{larger scale} than was
previously possible~(\S\ref{sec:platforms});
and
(iii)\,a methodology for conducting statistical inference of results generated in
scaled-down (sampled) Tor networks~(\S\ref{sec:significance}).
We showcase our modeling and simulation scalability improvements by
running simulations with 6,489 relays and 792k users, the largest
known Tor network simulations and the first at a network scale of
100\%~(\S\ref{sec:platforms:validate}).
Building upon the above foundations, we conduct a case study of the effects of
traffic load on client performance in the Tor network through a total of 420 Tor
simulations across three network scale factors and two traffic load factors
(\S\ref{sec:casestudy}).
Our case study demonstrates how to apply our methodologies for modeling Tor
networks and for conducting sound statistical inferences of results.

\paragraph{Conclusions}
We find that:
(i)\,significant reductions in RAM are possible by representing
multiple Tor users in each Tor client process
(\S\ref{sec:platforms:validate});
(ii)\,it is feasible to run 100\% Tor network simulations on
high-memory servers in a reasonable time (less than 2 weeks)
(\S\ref{sec:platforms:validate});
(iii)\,running multiple simulations in independent Tor networks is
necessary to draw statistically significant conclusions
(\S\ref{sec:significance}); and
(iv)\,fewer simulations are generally needed to achieve a desired CI precision
in networks of larger scale than in those of smaller scale
(\S\ref{sec:casestudy}).

\paragraph{Limitations and Future Work}
Although routers in Shadow drop packets when congested (using CoDel), we
describe in \S\ref{sec:inetmodel} that we do not model any additional artificial
packet loss. However, it is possible that packet loss or corruption rates are
higher in Tor than in Shadow (e.g., for mobile clients that are wirelessly
connected), and modeling this loss could improve realism. Future work should
consider developing a more realistic packet loss model that is, for example, based on
measurements of actual Tor clients and relays.

In \S\ref{sec:tormodel:stage} we describe that we compute some relay
characteristics (e.g., consensus weight, bandwidth rate and burst, location)
using the median value of those observed across all consensus and server
descriptors from the staging period. Similarly, in \S\ref{sec:tormodel:generate}
we describe that we select $m$ relays from those available by ``bucketing'' them
and choosing the relay with the median bandwidth capacity from each bucket.
These selection criteria may not capture the full variance in the relay
characteristics. Future work might consider alternative selection
strategies---such as randomly sampling the full observed distribution of each
characteristic, choosing based on occurrence count, or choosing uniformly at
random---and evaluate how such choices affect simulation accuracy.

Our traffic modeling approach in \S\ref{sec:tormodel:generate} allows us to
reduce the RAM required to run simulations by simulating $1/p$ users in each Tor
client process. This optimization yields the following implications. First, we
disable guards in our model because Tor does not currently support multiple
guard ``sessions'' on a given Tor client. Future work should consider either
implementing support for guard ``sessions'' in the Tor client, or otherwise
managing guard selection and circuit assignment through the Tor control port.
Second, simulating $1/p$ users on a Tor client results in ``clustering'' these
users in the city that was assigned to the client, resulting in lower location
diversity. Choosing values of $p$ closer to 1 would reduce this effect. Third,
setting $p<1$ reduces the total number of Tor clients and therefore the total
number of network descriptor fetches. Because these fetches occur infrequently
in Tor, the network impact is negligible relative to the total amount of traffic
being generated by each client.

Finally, future work might consider sampling the Tor network at scales $s>1.0$,
which could help us better understand how Tor might handle growth as it becomes
more popular.
	
\paragraph{Acknowledgments}
We thank our shepherd, Yixin Sun, and the anonymous reviewers for their valuable
feedback. This work has been partially supported by the Office of Naval Research
(ONR), the Defense Advanced Research Projects Agency (DARPA), the National
Science Foundation (NSF) under award CNS-1925497, and the
National Sciences and Engineering Research Council of Canada (NSERC) under award CRDPJ-534381.
This research was undertaken, in part, thanks to funding from the Canada Research Chairs program.
This work benefited from the use of the CrySP RIPPLE Facility at the University of Waterloo.

{\small
\bibliographystyle{abbrvnat}
\bibliography{references}

\begin{thebibliography}{75}
\providecommand{\natexlab}[1]{#1}
\providecommand{\url}[1]{\texttt{#1}}
\expandafter\ifx\csname urlstyle\endcsname\relax
  \providecommand{\doi}[1]{doi: #1}\else
  \providecommand{\doi}{doi: \begingroup \urlstyle{rm}\Url}\fi

\bibitem[Acquisti et~al.(2003)Acquisti, Dingledine, and Syverson]{econymics}
A.~Acquisti, R.~Dingledine, and P.~Syverson.
\newblock {On the Economics of Anonymity}.
\newblock In \emph{7th International Financial Cryptography Conference (FC)},
  2003.

\bibitem[AlSabah and Goldberg(2013)]{AlSabah2013pctcp}
M.~AlSabah and I.~Goldberg.
\newblock {PCTCP: Per-circuit TCP-over-IPsec Transport for Anonymous
  Communication Overlay Networks}.
\newblock In \emph{ACM Conference on Computer and Communications Security
  (CCS)}, 2013.

\bibitem[AlSabah and Goldberg(2016)]{alsabah2016performance}
M.~AlSabah and I.~Goldberg.
\newblock {Performance and Security Improvements for Tor: A Survey}.
\newblock \emph{ACM Computing Surveys (CSUR)}, 49\penalty0 (2):\penalty0 32,
  2016.

\bibitem[AlSabah et~al.(2011)AlSabah, Bauer, Goldberg, Grunwald, McCoy, Savage,
  and Voelker]{alsabah2011defenestrator}
M.~AlSabah, K.~Bauer, I.~Goldberg, D.~Grunwald, D.~McCoy, S.~Savage, and G.~M.
  Voelker.
\newblock {DefenestraTor: Throwing Out Windows in {Tor}}.
\newblock In \emph{Privacy Enhancing Technologies Symposium (PETS)}, pages
  134--154, 2011.

\bibitem[AlSabah et~al.(2013)AlSabah, Bauer, Elahi, and
  Goldberg]{alsabah2013path}
M.~AlSabah, K.~Bauer, T.~Elahi, and I.~Goldberg.
\newblock {The Path Less Travelled: Overcoming Tor's Bottlenecks with Traffic
  Splitting}.
\newblock In \emph{Privacy Enhancing Technologies Symposium (PETS)}, 2013.

\bibitem[Aryan et~al.(2013)Aryan, Aryan, and Halderman]{Aryan2013a}
S.~Aryan, H.~Aryan, and J.~A. Halderman.
\newblock {{Internet} Censorship in {Iran}: A First Look}.
\newblock In \emph{3rd USENIX Workshop on Free and Open Communications on the
  Internet (FOCI)}, 2013.

\bibitem[Barton and Wright(2016)]{barton2016denasa}
A.~Barton and M.~Wright.
\newblock {{DeNASA}: Destination-Naive {AS}-Awareness in Anonymous
  Communications}.
\newblock \emph{Proceedings on Privacy Enhancing Technologies (PoPETs)},
  2016\penalty0 (4):\penalty0 356--372, 2016.

\bibitem[Bauer et~al.(2011)Bauer, Sherr, and Grunwald]{bauer2011experimentor}
K.~S. Bauer, M.~Sherr, and D.~Grunwald.
\newblock {{ExperimenTor}: A Testbed for Safe and Realistic {Tor}
  Experimentation}.
\newblock In \emph{USENIX Workshop on Cyber Security Experimentation and Test
  (CSET)}, 2011.

\bibitem[Blumofe and Leiserson(1999)]{worksteal}
R.~D. Blumofe and C.~E. Leiserson.
\newblock {Scheduling Multithreaded Computations by Work Stealing}.
\newblock \emph{J. ACM}, 46\penalty0 (5):\penalty0 720--748, Sept. 1999.

\bibitem[Brave(2019)]{bravesite}
Brave.
\newblock {Brave Browser}.
\newblock \url{https://brave.com/}, November 2019.
\newblock Accessed 2020-09-30.

\bibitem[Clark et~al.(2007)Clark, van Oorschot, and Adams]{clark2007usability}
J.~Clark, P.~C. van Oorschot, and C.~Adams.
\newblock {Usability of Anonymous Web Browsing: An Examination of Tor
  Interfaces and Deployability}.
\newblock In \emph{{3rd Symposium on Usable Privacy and Security (SOUPS)}},
  2007.

\bibitem[Conrad and Shirazi(2014)]{conrad2014analyzing}
B.~Conrad and F.~Shirazi.
\newblock {Analyzing the Effectiveness of DoS Attacks on Tor}.
\newblock In \emph{7th International Conference on Security of Information and
  Networks}, page 355, 2014.

\bibitem[Dahal et~al.(2015)Dahal, Lee, Kang, and Shin]{icoin15}
S.~Dahal, J.~Lee, J.~Kang, and S.~Shin.
\newblock {Analysis on End-to-End Node Selection Probability in Tor Network}.
\newblock In \emph{2015 International Conference on Information Networking
  (ICOIN)}, pages 46--50, Jan 2015.

\bibitem[Dingledine and Mathewson(2006)]{usability:weis2006}
R.~Dingledine and N.~Mathewson.
\newblock {Anonymity Loves Company: Usability and the Network Effect}.
\newblock In \emph{5th Workshop on the Economics of Information Security
  (WEIS)}, 2006.

\bibitem[Dingledine et~al.(2004)Dingledine, Mathewson, and
  Syverson]{tor-design}
R.~Dingledine, N.~Mathewson, and P.~Syverson.
\newblock {{Tor}: The Second-Generation Onion Router}.
\newblock In \emph{USENIX Security Symposium (USENIX-Sec)}, 2004.

\bibitem[Dinh et~al.(2020)Dinh, Rochet, Pereira, and Wallach]{dinh2020scaling}
T.-N. Dinh, F.~Rochet, O.~Pereira, and D.~S. Wallach.
\newblock {Scaling Up Anonymous Communication with Efficient Nanopayment
  Channels}.
\newblock \emph{Proceedings on Privacy Enhancing Technologies (PoPETs)},
  2020\penalty0 (3):\penalty0 175--203, 2020.

\bibitem[Elahi et~al.(2014)Elahi, Danezis, and Goldberg]{PrivExElahi2014}
T.~Elahi, G.~Danezis, and I.~Goldberg.
\newblock {{PrivEx}: Private Collection of Traffic Statistics for Anonymous
  Communication Networks}.
\newblock In \emph{ACM Conference on Computer and Communications Security
  (CCS)}, 2014.
\newblock See also \url{git://git-crysp.uwaterloo.ca/privex}.

\bibitem[Fabian et~al.(2010)Fabian, Goertz, Kunz, M{\"u}ller, and
  Nitzsche]{privatelywaiting}
B.~Fabian, F.~Goertz, S.~Kunz, S.~M{\"u}ller, and M.~Nitzsche.
\newblock {Privately Waiting --- A Usability Analysis of the Tor Anonymity
  Network}.
\newblock In \emph{Sustainable e-Business Management}, 2010.

\bibitem[Fenske et~al.(2017)Fenske, Mani, Johnson, and
  Sherr]{fenske2017distributed}
E.~Fenske, A.~Mani, A.~Johnson, and M.~Sherr.
\newblock {Distributed Measurement with Private Set-Union Cardinality}.
\newblock In \emph{ACM Conference on Computer and Communications Security
  (CCS)}, 2017.

\bibitem[Geddes et~al.(2013)Geddes, Jansen, and Hopper]{geddes2013low}
J.~Geddes, R.~Jansen, and N.~Hopper.
\newblock {How Low Can You Go: Balancing Performance with Anonymity in Tor}.
\newblock In \emph{13th Privacy Enhancing Technologies Symposium}, pages
  164--184, 2013.

\bibitem[Geddes et~al.(2014)Geddes, Jansen, and Hopper]{geddes2014imux}
J.~Geddes, R.~Jansen, and N.~Hopper.
\newblock {{IMUX}: Managing {Tor} Connections from Two to Infinity, and
  Beyond}.
\newblock In \emph{ACM Workshop on Privacy in the Electronic Society (WPES)},
  pages 181--190, 2014.

\bibitem[Geddes et~al.(2016)Geddes, Schliep, and Hopper]{geddes2016abra}
J.~Geddes, M.~Schliep, and N.~Hopper.
\newblock {ABRA CADABRA: Magically Increasing Network Utilization in Tor by
  Avoiding Bottlenecks}.
\newblock In \emph{15th ACM Workshop on Privacy in the Electronic Society},
  pages 165--176, 2016.

\bibitem[Gopal and Heninger(2012)]{gopal2012torchestra}
D.~Gopal and N.~Heninger.
\newblock {Torchestra: Reducing Interactive Traffic Delays over {Tor}}.
\newblock In \emph{ACM Workshop on Privacy in the Electronic Society (WPES)},
  2012.

\bibitem[Hanley et~al.(2019)Hanley, Sun, Wagh, and Mittal]{hanley2019dpselect}
H.~Hanley, Y.~Sun, S.~Wagh, and P.~Mittal.
\newblock {DPSelect: A Differential Privacy Based Guard Relay Selection
  Algorithm for Tor}.
\newblock \emph{Proceedings on Privacy Enhancing Technologies (PoPETs)},
  2019\penalty0 (2):\penalty0 166--186, 2019.

\bibitem[Hoel(1971)]{statsbook}
P.~G. Hoel.
\newblock \emph{{Introduction to Mathematical Statistics}}.
\newblock Wiley, New York, 4th edition, 1971.
\newblock ISBN 0471403652.

\bibitem[Hopper(2014)]{hopper2014challenges}
N.~Hopper.
\newblock {Challenges in protecting Tor hidden services from botnet abuse}.
\newblock In \emph{Financial Cryptography and Data Security (FC)}, pages
  316--325, 2014.

\bibitem[Imani et~al.(2018)Imani, Barton, and Wright]{imani2018guard}
M.~Imani, A.~Barton, and M.~Wright.
\newblock {Guard Sets in Tor using AS Relationships}.
\newblock \emph{Proceedings on Privacy Enhancing Technologies (PoPETs)},
  2018\penalty0 (1):\penalty0 145--165, 2018.

\bibitem[Imani et~al.(2019)Imani, Amirabadi, and Wright]{imani2019modified}
M.~Imani, M.~Amirabadi, and M.~Wright.
\newblock {Modified Relay Selection and Circuit Selection for Faster Tor}.
\newblock \emph{IET Communications}, 13\penalty0 (17):\penalty0 2723--2734,
  2019.

\bibitem[Jansen and Hopper(2012)]{jansen2012shadow}
R.~Jansen and N.~Hopper.
\newblock {{Shadow}: Running {Tor} in a Box for Accurate and Efficient
  Experimentation}.
\newblock In \emph{Network and Distributed System Security Symposium (NDSS)},
  2012.
\newblock See also \url{https://shadow.github.io}.

\bibitem[Jansen and Johnson(2016)]{privcount-ccs2016}
R.~Jansen and A.~Johnson.
\newblock {Safely Measuring {Tor}}.
\newblock In \emph{ACM Conference on Computer and Communications Security
  (CCS)}, 2016.
\newblock See also \url{https://github.com/privcount}.

\bibitem[Jansen et~al.(2010)Jansen, Hopper, and Kim]{jansen2010recruiting}
R.~Jansen, N.~Hopper, and Y.~Kim.
\newblock {Recruiting New {Tor} Relays with {BRAIDS}}.
\newblock In \emph{ACM Conference on Computer and Communications Security
  (CCS)}, 2010.

\bibitem[Jansen et~al.(2012{\natexlab{a}})Jansen, Bauer, Hopper, and
  Dingledine]{jansen2012cset}
R.~Jansen, K.~Bauer, N.~Hopper, and R.~Dingledine.
\newblock {Methodically Modeling the {Tor} Network}.
\newblock In \emph{USENIX Workshop on Cyber Security Experimentation and Test
  (CSET)}, 2012{\natexlab{a}}.

\bibitem[Jansen et~al.(2012{\natexlab{b}})Jansen, Syverson, and
  Hopper]{jansen2012throttling}
R.~Jansen, P.~F. Syverson, and N.~Hopper.
\newblock {Throttling {Tor} Bandwidth Parasites}.
\newblock In \emph{USENIX Security Symposium (USENIX-Sec)}, 2012{\natexlab{b}}.

\bibitem[Jansen et~al.(2013)Jansen, Johnson, and Syverson]{jansen2013lira}
R.~Jansen, A.~Johnson, and P.~Syverson.
\newblock {LIRA: Lightweight Incentivized Routing for Anonymity}.
\newblock In \emph{Network and Distributed System Security Symposium (NDSS)},
  2013.

\bibitem[Jansen et~al.(2014{\natexlab{a}})Jansen, Geddes, Wacek, Sherr, and
  Syverson]{jansen2014kist}
R.~Jansen, J.~Geddes, C.~Wacek, M.~Sherr, and P.~Syverson.
\newblock {Never Been {KIST}: {Tor}'s Congestion Management Blossoms with
  Kernel-Informed Socket Transport}.
\newblock In \emph{USENIX Security Symposium (USENIX-Sec)}, 2014{\natexlab{a}}.

\bibitem[Jansen et~al.(2014{\natexlab{b}})Jansen, Tschorsch, Johnson, and
  Scheuermann]{jansen2014sniper}
R.~Jansen, F.~Tschorsch, A.~Johnson, and B.~Scheuermann.
\newblock {The Sniper Attack: Anonymously Deanonymizing and Disabling the {Tor}
  Network}.
\newblock In \emph{Network and Distributed System Security Symposium (NDSS)},
  2014{\natexlab{b}}.

\bibitem[Jansen et~al.(2018{\natexlab{a}})Jansen, Traudt, Geddes, Wacek, Sherr,
  and Syverson]{kist-tops2018}
R.~Jansen, M.~Traudt, J.~Geddes, C.~Wacek, M.~Sherr, and P.~Syverson.
\newblock {KIST: Kernel-Informed Socket Transport for {Tor}}.
\newblock \emph{ACM Transactions on Privacy and Security (TOPS)}, 22\penalty0
  (1):\penalty0 3:1--3:37, December 2018{\natexlab{a}}.

\bibitem[Jansen et~al.(2018{\natexlab{b}})Jansen, Traudt, and
  Hopper]{tmodel-ccs2018}
R.~Jansen, M.~Traudt, and N.~Hopper.
\newblock {Privacy-Preserving Dynamic Learning of {Tor} Network Traffic}.
\newblock In \emph{ACM Conference on Computer and Communications Security
  (CCS)}, 2018{\natexlab{b}}.
\newblock See also \url{https://tmodel-ccs2018.github.io}.

\bibitem[Jansen et~al.(2019)Jansen, Vaidya, and Sherr]{jansen2019point}
R.~Jansen, T.~Vaidya, and M.~Sherr.
\newblock {Point Break: A Study of Bandwidth Denial-of-Service Attacks against
  {Tor}}.
\newblock In \emph{USENIX Security Symposium (USENIX-Sec)}, 2019.

\bibitem[Johnson et~al.(2017{\natexlab{a}})Johnson, Jansen, Hopper, Segal, and
  Syverson]{johnson2017peerflow}
A.~Johnson, R.~Jansen, N.~Hopper, A.~Segal, and P.~Syverson.
\newblock {{PeerFlow}: Secure Load Balancing in {Tor}}.
\newblock \emph{Proceedings on Privacy Enhancing Technologies (PoPETs)},
  2017\penalty0 (2):\penalty0 74--94, 2017{\natexlab{a}}.

\bibitem[Johnson et~al.(2017{\natexlab{b}})Johnson, Jansen, Jaggard,
  Feigenbaum, and Syverson]{johnson2017avoiding}
A.~Johnson, R.~Jansen, A.~D. Jaggard, J.~Feigenbaum, and P.~Syverson.
\newblock {Avoiding The Man on the Wire: Improving {Tor}'s Security with
  Trust-Aware Path Selection}.
\newblock In \emph{Network and Distributed System Security Symposium (NDSS)},
  2017{\natexlab{b}}.

\bibitem[Kiran et~al.(2019)Kiran, Chalke, Usman, Shenoy, and
  Venugopal]{kiran2019anonymity}
K.~Kiran, S.~S. Chalke, M.~Usman, P.~D. Shenoy, and K.~Venugopal.
\newblock {Anonymity and Performance Analysis of Stream Isolation in Tor
  Network}.
\newblock In \emph{International Conference on Computing, Communication and
  Networking Technologies (ICCCNT)}, 2019.

\bibitem[Komlo et~al.(2020)Komlo, Mathewson, and Goldberg]{walkingonions}
C.~H. Komlo, N.~Mathewson, and I.~Goldberg.
\newblock Walking onions: Scaling anonymity networks while protecting users.
\newblock In \emph{USENIX Security Symposium (USENIX-Sec)}, 2020.

\bibitem[Lakshmikantha et~al.(2008)Lakshmikantha, Srikant, and
  Beck]{corerouters}
A.~Lakshmikantha, R.~Srikant, and C.~Beck.
\newblock {Impact of File Arrivals and Departures on Buffer Sizing in Core
  Routers}.
\newblock In \emph{IEEE INFOCOM 2008 - The 27th Conference on Computer
  Communications}, May 2008.

\bibitem[Lee et~al.(2017)Lee, Fifield, Malkin, Iyer, Egelman, and
  Wagner]{lee2017usability}
L.~Lee, D.~Fifield, N.~Malkin, G.~Iyer, S.~Egelman, and D.~Wagner.
\newblock {A Usability Evaluation of Tor Launcher}.
\newblock \emph{Proceedings on Privacy Enhancing Technologies (PoPETs)}, 2017,
  07 2017.

\bibitem[Lin et~al.(2016)Lin, Sherr, and Loo]{lin2016scalable}
D.~Lin, M.~Sherr, and B.~T. Loo.
\newblock {Scalable and Anonymous Group Communication with {MTor}}.
\newblock \emph{Proceedings on Privacy Enhancing Technologies (PoPETs)},
  2016\penalty0 (2):\penalty0 22--39, 2016.

\bibitem[Liu et~al.(2017)Liu, Liu, Winter, Mittal, and Hu]{Liu2017a}
Z.~Liu, Y.~Liu, P.~Winter, P.~Mittal, and Y.-C. Hu.
\newblock {{TorPolice}: Towards Enforcing Service-Defined Access Policies for
  Anonymous Communication in the {Tor} Network}.
\newblock In \emph{International Conference on Network Protocols}, 2017.

\bibitem[Loesing et~al.(2010)Loesing, Murdoch, and
  Dingledine]{loesing2010measuring}
K.~Loesing, S.~J. Murdoch, and R.~Dingledine.
\newblock {A Case Study on Measuring Statistical Data in the {Tor} Anonymity
  Network}.
\newblock In \emph{Financial Cryptography and Data Security (FC)}, 2010.
\newblock See also \url{https://metrics.torproject.org}.

\bibitem[Mani and Sherr(2017)]{mani2017historvarepsilon}
A.~Mani and M.~Sherr.
\newblock {{HisTor$\varepsilon$}: Differentially Private and Robust Statistics
  Collection for {Tor}}.
\newblock In \emph{Network and Distributed System Security Symposium (NDSS)},
  2017.

\bibitem[Mani et~al.(2018)Mani, Wilson-Brown, Jansen, Johnson, and
  Sherr]{torusage-imc2018}
A.~Mani, T.~Wilson-Brown, R.~Jansen, A.~Johnson, and M.~Sherr.
\newblock {Understanding {Tor} Usage with Privacy-Preserving Measurement}.
\newblock In \emph{18th ACM Internet Measurement Conference (IMC)}, 2018.
\newblock See also \url{https://torusage-imc2018.github.io}.

\bibitem[Miller and Jansen(2015)]{miller2015shadow}
A.~Miller and R.~Jansen.
\newblock {{Shadow-Bitcoin}: Scalable Simulation via Direct Execution of
  Multi-threaded Applications}.
\newblock In \emph{USENIX Workshop on Cyber Security Experimentation and Test
  (CSET)}, 2015.

\bibitem[Mitseva et~al.(2020)Mitseva, Aleksandrova, Engel, and
  Panchenko]{mitseva2020security}
A.~Mitseva, M.~Aleksandrova, T.~Engel, and A.~Panchenko.
\newblock {Security and Performance Implications of BGP Rerouting-Resistant
  Guard Selection Algorithms for Tor}.
\newblock In \emph{IFIP International Conference on ICT Systems Security and
  Privacy Protection}, 2020.

\bibitem[Moore et~al.(2011)Moore, Wacek, and Sherr]{Moore2011tortoise}
W.~B. Moore, C.~Wacek, and M.~Sherr.
\newblock {Exploring the Potential Benefits of Expanded Rate Limiting in {Tor}:
  Slow and Steady Wins the Race with Tortoise}.
\newblock In \emph{Annual Computer Security Applications Conference (ACSAC)},
  2011.

\bibitem[Mozilla(2019)]{mozillacallcfp}
Mozilla.
\newblock {Mozilla Research Grants 2019H1}.
\newblock
  \url{https://mozilla-research.forms.fm/mozilla-research-grants-2019h1/forms/6510},
  2019.
\newblock Call for Proposals.

\bibitem[{Mozilla}(2019)]{mozillascale}
{Mozilla}.
\newblock {Firefox Public Data Report}.
\newblock \url{https://data.firefox.com/dashboard/user-activity}, December
  2019.

\bibitem[Ngan et~al.(2010)Ngan, Dingledine, and Wallach]{incentives-fc10}
T.-W.~J. Ngan, R.~Dingledine, and D.~S. Wallach.
\newblock {Building Incentives into {Tor}}.
\newblock In \emph{Financial Cryptography and Data Security (FC)}, 2010.

\bibitem[Norcie et~al.(2012)Norcie, Caine, and Camp]{norcie2012stoppoints}
G.~Norcie, K.~Caine, and L.~J. Camp.
\newblock {Eliminating Stop-Points in the Installation and Use of Anonymity
  Systems: a Usability Evaluation of the Tor Browser Bundle}.
\newblock In \emph{Privacy Enhancing Technologies Symposium (PETS)}, 2012.

\bibitem[Rochet and Pereira(2017)]{rochet2017waterfilling}
F.~Rochet and O.~Pereira.
\newblock {Waterfilling: Balancing the Tor network with maximum diversity}.
\newblock \emph{Proceedings on Privacy Enhancing Technologies (PoPETs)},
  2017\penalty0 (2):\penalty0 4--22, 2017.

\bibitem[Rochet and Pereira(2018)]{rochet2018dropping}
F.~Rochet and O.~Pereira.
\newblock {Dropping on the Edge: Flexibility and Traffic Confirmation in Onion
  Routing Protocols}.
\newblock \emph{Proceedings on Privacy Enhancing Technologies (PoPETs)},
  2018\penalty0 (2):\penalty0 27--46, 2018.

\bibitem[Rochet et~al.(2020)Rochet, Wails, Johnson, Mittal, and
  Pereira]{rochet2020claps}
F.~Rochet, R.~Wails, A.~Johnson, P.~Mittal, and O.~Pereira.
\newblock {CLAPS: Client-Location-Aware Path Selection in Tor}.
\newblock In \emph{ACM Conference on Computer and Communications Security
  (CCS)}, 2020.

\bibitem[Shirazi et~al.(2015{\natexlab{a}})Shirazi, Diaz, and
  Wright]{shirazi2015towards}
F.~Shirazi, C.~Diaz, and J.~Wright.
\newblock {Towards Measuring Resilience in Anonymous Communication Networks}.
\newblock In \emph{14th ACM Workshop on Privacy in the Electronic Society},
  pages 95--99, 2015{\natexlab{a}}.

\bibitem[Shirazi et~al.(2015{\natexlab{b}})Shirazi, Goehring, and
  Diaz]{torexptools}
F.~Shirazi, M.~Goehring, and C.~Diaz.
\newblock {Tor Experimentation Tools}.
\newblock In \emph{International Workshop on Privacy Engineering (IWPE)},
  2015{\natexlab{b}}.

\bibitem[Singh(2014)]{singh2014thesis}
S.~Singh.
\newblock {Large-Scale Emulation of Anonymous Communication Networks}.
\newblock Master's thesis, University of Waterloo, 2014.

\bibitem[Tang and Goldberg(2010)]{ccs10-scheduling}
C.~Tang and I.~Goldberg.
\newblock {An Improved Algorithm for {T}or Circuit Scheduling}.
\newblock In \emph{17th ACM Conference on Computer and Communications Security
  (CCS)}, 2010.

\bibitem[The Tor Project()]{mozillacallblog}
The Tor Project.
\newblock {Mozilla Research Call: Tune up Tor for Integration and Scale}.
\newblock
  \url{https://blog.torproject.org/mozilla-research-call-tune-tor-integration-and-scale},
  May 2019.
\newblock Blog Post.

\bibitem[{The Tor Project}(2020{\natexlab{a}})]{tormetrics}
{The Tor Project}.
\newblock {Tor Metrics Portal}.
\newblock \url{https://metrics.torproject.org}, January 2020{\natexlab{a}}.

\bibitem[{The Tor Project}(2020{\natexlab{b}})]{tormetrics-repro}
{The Tor Project}.
\newblock {Reproducible Metrics}.
\newblock
  \url{https://metrics.torproject.org/reproducible-metrics.html#performance},
  October 2020{\natexlab{b}}.

\bibitem[{The Tor Project}(2020{\natexlab{c}})]{torproject}
{The Tor Project}.
\newblock {The Tor Project}.
\newblock \url{https://www.torproject.org}, January 2020{\natexlab{c}}.

\bibitem[Tracey et~al.(2018)Tracey, Jansen, and Goldberg]{tracey2018elfs}
J.~Tracey, R.~Jansen, and I.~Goldberg.
\newblock {High Performance {Tor} Experimentation from the Magic of Dynamic
  {ELF}s}.
\newblock In \emph{USENIX Workshop on Cyber Security Experimentation and Test
  (CSET)}, 2018.

\bibitem[Unger(2018)]{netmirage}
N.~Unger.
\newblock {NetMirage}.
\newblock \url{https://crysp.uwaterloo.ca/software/netmirage/}, 2018.
\newblock Accessed 2020-02-12.

\bibitem[{United Nations}(2020)]{freeinfo}
{United Nations}.
\newblock {Freedom of Information}.
\newblock
  \url{https://www.un.org/ruleoflaw/thematic-areas/governance/freedom-of-information},
  January 2020.

\bibitem[Vahdat et~al.(2003)Vahdat, Yocum, Walsh, Mahadevan, Kosti\'{c}, Chase,
  and Becker]{modelnet}
A.~Vahdat, K.~Yocum, K.~Walsh, P.~Mahadevan, D.~Kosti\'{c}, J.~Chase, and
  D.~Becker.
\newblock {Scalability and Accuracy in a Large-Scale Network Emulator}.
\newblock \emph{SIGOPS Oper. Syst. Rev.}, 36\penalty0 (SI):\penalty0 271--284,
  Dec. 2003.

\bibitem[Wacek et~al.(2013)Wacek, Tan, Bauer, and Sherr]{tor-relayselection}
C.~Wacek, H.~Tan, K.~Bauer, and M.~Sherr.
\newblock {An Empirical Evaluation of Relay Selection in {Tor}}.
\newblock In \emph{Network and Distributed System Security Symposium (NDSS)},
  2013.

\bibitem[Yang and Li(2015{\natexlab{a}})]{mtor-cns15}
L.~Yang and F.~Li.
\newblock {mTor: A Multipath Tor Routing Beyond Bandwidth Throttling}.
\newblock In \emph{2015 IEEE Conference on Communications and Network Security
  (CNS)}, pages 479--487, Sept 2015{\natexlab{a}}.

\bibitem[Yang and Li(2015{\natexlab{b}})]{yang2015enhancing}
L.~Yang and F.~Li.
\newblock {Enhancing Traffic Analysis Resistance for Tor Hidden Services with
  Multipath Routing}.
\newblock In \emph{International Conference on Security and Privacy in
  Communication Systems}, pages 367--384, 2015{\natexlab{b}}.

\end{thebibliography}
}

\vspace{-2mm}
\appendix

\section*{Appendix}

\section{Ontology of Tor Performance Metrics} \label{sec:ontol}
\newcommand{\subarrow}[1][]{
  \begin{tikzpicture}[#1]
    \draw[semithick] (0,0.7ex) -- (0,0) -- (0.75em,0);
    \draw[thick] (0.55em,0.2em) -- (0.75em,0) -- (0.55em,-0.2em);
  \end{tikzpicture}
}
\newcommand\subproperty[1]{\subarrow~{#1}}

\newcommand{\ssubarrow}[1][]{
  \!\!\!\!\!\!
  \begin{tikzpicture}[#1]
    \draw[semithick] (-0.7ex,0.7ex) -- (0,0) -- (0.75em,0);
    \draw[thick] (0.55em,0.2em) -- (0.75em,0) -- (0.55em,-0.2em);
  \end{tikzpicture}
}
\newcommand\ssubproperty[1]{\ssubarrow~{#1}}

\renewcommand*\rothead[1]{\hspace{3.42mm}\begin{rotate}{45}#1\end{rotate}}
\begin{table*}[p]
  \centering
  \footnotesize
  \caption{Classification of the experimentation properties in our Tor ontology into Independent and Dependent variables, organized by element.     An arrow indicates a subproperty.  }
  \label{tab:ontology}
    \begin{tabular}{l|RRRRRRRRR@{}|@{}RRRRRRRRRRR@{}|@{}RRRRRRRRRRRR@{}|@{}RRRRRRRRRR@{}|RR|}
    \multicolumn{1}{c}{} \\[9ex]     \multicolumn{1}{c}{}
    & \rothead{latency}
    & \rothead{\ssubproperty{jitter}}
    & \rothead{bandwidth}
    & \rothead{reliability}
    & \rothead{\ssubproperty{packet loss}}
    & \rothead{path/routing}
    & \rothead{congestion}
    & \rothead{time to Tor consensus}
    & \multicolumn{1}{D}{\rothead{time to steady state}}
        & \rothead{quantity}
    & \rothead{IP address}
    & \rothead{geolocation}
    & \rothead{stack}
    & \rothead{\ssubproperty{OS/kernel}}
    & \rothead{\ssubproperty{hardware}}
    & \rothead{CPU/memory usage}
    & \rothead{throughput}
    & \rothead{goodput}
    & \rothead{control overhead}
    & \multicolumn{1}{D}{\rothead{\ssubproperty{retransmissions}}}
        & \rothead{behavior model}
    & \rothead{\ssubproperty{number of connections}}
    & \rothead{\ssubproperty{duration of connections}}
    & \rothead{\ssubproperty{traffic type}}
    & \rothead{\ssubproperty{idle time}}
    & \rothead{time to first/last byte}
    & \rothead{Tor}
    & \rothead{\ssubproperty{relay selection algorithm}}
    & \rothead{\ssubproperty{max num. open circuits}}
    & \rothead{\ssubproperty{max duration of circuits}}
    & \rothead{\ssubproperty{circuit build time}}
    & \multicolumn{1}{D}{\rothead{\ssubproperty{errors}}}
        & \rothead{Tor relay configuration}
    & \rothead{packet delay}
    & \rothead{congestion}
    & \rothead{processing (Tor)}
    & \rothead{processing (stack)}
    & \rothead{connections (number of:)}
    & \rothead{\ssubproperty{sockets open}}
    & \rothead{\ssubproperty{streams and circuits}}
    & \rothead{\ssubproperty{connecting clients}}
    & \multicolumn{1}{S}{\rothead{\ssubproperty{errors}}}
        & \rothead{behavior model}
    & \multicolumn{1}{R}{\rothead{port}}
    \\
    \cline{2-45}
    Indep. & \yes{} & \yes{} & \yes{} & \yes{} & \yes{} & \yes{} & \no{}  & \no{}  & \no{}
    & \yes{} & \yes{} & \yes{} & \yes{} & \yes{} & \yes{} & \no{}  & \no{}  & \no{}  & \no{}  & \no{}
    & \yes{} & \yes{} & \yes{} & \yes{} & \yes{} & \no{}  & & \yes{} & \yes{} & \yes{} & \no{} & \no{}
    & \yes{} & \no{}  & \no{}  & \no{}  & \no{}  & & \no{}  & \no{}  & \no{}  & \no{}
    & \yes{} & \yes{} \\
    Dep.   & \no{}  & \no{}  & \no{}  & \yes{} & \yes{} & \yes{} & \yes{} & \yes{} & \yes{}
    & \no{}  & \no{}  & \no{}  & \no{}  & \no{}  & \no{}  & \yes{} & \yes{} & \yes{} & \yes{} & \yes{}
    & \no{}  & \no{}  & \no{}  & \no{}  & \no{}  & \yes{} & & \no{}  & \yes{} & \yes{} & \yes{} & \yes{}
    & \no{}  & \yes{} & \yes{} & \yes{} & \yes{} & & \yes{} & \yes{} & \yes{} & \yes{}
    & \no{}  & \no{}  \\
    \cline{2-45}
    \multicolumn{1}{c|}{} & \multicolumn{9}{c|}{\multirow{2}{*}{Network}}
    & \multicolumn{11}{c|}{Common} & \multicolumn{12}{c|}{Clients} & \multicolumn{10}{c|}{Relays} & \multicolumn{2}{c|}{Servers} \\
    \cline{11-45}
    \multicolumn{1}{c|}{} & \multicolumn{9}{c|}{} & \multicolumn{35}{c|}{Network Nodes} \\
    \cline{2-45}
\end{tabular}~~~~~
\end{table*}

\begin{table*}[p]
\centering
\footnotesize
\caption{Description of properties from the ontology.
Arrows denote subproperties.}
\label{tab:ontol:descript}
\begin{tabularx}{\textwidth}{|l|l|lX}
\toprule
\multicolumn{2}{l}{} & \textbf{Property} & \textbf{Description} \\
\cline{1-4}\multicolumn{2}{|c|}{}&\\[-2ex]
  \multicolumn{2}{|c|}{\multirow{9}{*}{\begin{sideways}\textbf{Network}\end{sideways}}}
  & latency & The amount of time it takes for a packet to traverse from one network node to another.\\
  \multicolumn{2}{|c|}{} & \subproperty{jitter} & The variation in latency.\\
  \multicolumn{2}{|c|}{} & bandwidth & The amount of data a network connection can transfer in a given amount of time.\\
  \multicolumn{2}{|c|}{} & reliability & The probability of a network connection successfully transferring data.\\
  \multicolumn{2}{|c|}{} & \subproperty{packet loss} & The probability of a packet on an existing connection not arriving at the destination.\\
  \multicolumn{2}{|c|}{} & path/routing & The set of network nodes a packet passes through to arrive at its destination.\\
  \multicolumn{2}{|c|}{} & congestion & The amount of traffic load exceeding the capacity of the link or network node.\\
  \multicolumn{2}{|c|}{} & time to Tor consensus & The amount of time until the Tor network generates a valid consensus file (the file directory authorities publish containing information about every relay).\\
  \multicolumn{2}{|c|}{} & time to steady state & The amount of time until the network displays consistent behavior.\\
\cline{1-4}&\\[-2ex]
  \multirow{33}{*}{\begin{sideways}\textbf{Network Nodes}\end{sideways}}
  & \multirow{11}{*}{\begin{sideways}\textbf{Common}\end{sideways}}
  & quantity & The amount of this particular type of node in the network.\\
  & & IP address & The external IP address of the node.\\
  & & geolocation & Where the node is geographically located.\\
  & & stack & What Tor and associated processes are running on.\\
  & & \subproperty{OS/kernel} & The operating system, especially the network stack.\\
  & & \subproperty{hardware} & The computer components and their characteristics, such as CPU speed and memory capacity.\\
  & & CPU/memory usage & The amount of CPU time and RAM used.\\
  & & throughput & The total network traffic seen in a given amount of time, including overhead such as packet headers and retransmissions.\\
  & & goodput & The total amount of usable traffic seen in a given amount of time, therefore not including overhead from headers or retransmissions.\\
  & & control overhead & The amount of traffic that is spent on protocol data, rather than payload data.\\
  & & \subproperty{retransmissions} & The amount of traffic that was duplicated as a result of TCP acknowledgements not being received (in time).\\
\cline{2-4}&\\[-2ex]
  & \multirow{10}{*}{\begin{sideways}\textbf{Clients}\end{sideways}}
  & behavior model & How the client behaves.\\
  & & \subproperty{number of connections} & How many network connections the client creates (typically to servers, via Tor relays).\\
  & & \subproperty{duration of connections} & How long network connections last before being closed.\\
  & & \subproperty{traffic type} & The protocol and traffic properties (e.g., web pages, large downloads).\\
  & & \subproperty{idle time} & The time spent not sending any traffic, either because there is nothing being sent over a currently active connection, or because the client has completed all connections and has not yet started another.\\
  & & time to first byte & The amount of time it takes to receive the first byte of a download. Also known as round trip time (RTT).\\
  & & time to last byte & The amount of time it takes to complete a download.\\
  & & Tor & \\
  & & \subproperty{relay selection algorithm} & How Tor chooses which relays to route through.\\
  & & \subproperty{max number of open circuits} & The maximum number of Tor circuits simultaneously open.\\
  & & \subproperty{max duration of circuits} & The maximum amount of time circuits remain open.\\
  & & \subproperty{circuit build time} & How long it takes to construct a circuit.\\
  & & \subproperty{errors} & The number and characteristics of errors encountered.\\
\cline{2-4}&\\[-2ex]
  & \multirow{10}{*}{\begin{sideways}\textbf{Relays}\end{sideways}}
  & Tor relay configuration & The configuration of the Tor relay, whether in configuration files or changes to the Tor application. \\
  & & packet delay & The amount of additional time it takes for a packet to enter and leave the entire relay.\\
  & & congestion & Network congestion specifically as a result of the Tor relay process.\\
  & & processing (Tor) & The amount of time spent processing packets within the Tor process.\\
  & & processing (stack) & The amount of time spent processing packets outside the Tor process (primarily the OS).\\
  & & connections (number of:) & \\
  & & \subproperty{sockets open} & The number of network sockets the relay has open.\\
  & & \subproperty{streams and circuits} & The number of TCP streams and Tor circuits.\\
  & & \subproperty{connecting clients} & The number of clients that connect to this relay.\\
  & & \subproperty{errors} & The  number and characteristics of errors encountered.\\
\cline{2-4}&\\[-2ex]
  & \multirow{2}{*}{\begin{sideways}\textbf{Servers}\end{sideways}}
  & behavior model & How the server interacts with the client application communicating with it.\\
  & & port & The network ports the server is listening on. This is distinct from the behavior model in that Tor relays interact with it (via exit policies), not just the client.\\
\cline{1-4}
\end{tabularx}
\end{table*}

In this appendix, we describe an ontology of the Tor network, from the perspective and for the purpose of controlled performance research.
While our ontology shows one way of orienting known factors to consider when conducting Tor experiments,
we emphasize that it is not intended to be complete.
The most interesting future research may come not from the measurement of properties (or even elements) listed here, but from the gaps and undervalued areas
that are currently unexplored in Tor research.

\paragraph{Ontology}
The ontology consists of elements (clients, relays, servers, and the network), each of which have properties that can be further recursively subdivided into (sub)properties.
These properties can be viewed as the variables of an experiment, and therefore can be separated into independent and dependent variables.
Independent variables are properties that are \emph{set}, chosen during the course of experiment configuration (e.g., the number of clients, or available bandwidth).
Dependent variables are properties that can be \emph{measured} as results of the experiment (e.g., throughput, or successful downloads).
The division between what constitutes independent and dependent variables depends on the specific context of an experiment.
In this ontology, we classified properties based on the experimentation platforms we examined.
Specifically, these categorizations are based on controlled Tor experiments; more observational research (e.g., the measurements done on Tor Metrics~\cite{tormetrics}) would have a different perspective, primarily manifesting as many properties shifting from independent to dependent variables.
Even with this particular point of reference, however, some properties can concurrently exist as both independent and dependent variables in one experiment.
Packet loss, for example, is something that can be configured as a property of the network (i.e., a particular link can be configured to drop packets with some probability), but will also occur as a result of the natural behavior of TCP stacks establishing a stable connection and can therefore be measured.

The rest of this section is dedicated to describing the elements of our ontology. The properties of these elements are enumerated and classified in Table~\ref{tab:ontology}.
While most of the terms are self-explanatory, they are also briefly described in Table~\ref{tab:ontol:descript} to alleviate any confusion.

\paragraph{Network}
The network element represents the connections between other elements, as well as meta-properties that are not directly measurable on individual nodes (though analogues may be).
Latency and bandwidth, for example, are properties directly instantiated in the links between the other elements.
The time to a steady state, on the other hand, is something that can be measured, but not as an actual property of any particular element, so much as a measurement of a constructed
representation of the network itself.

\paragraph{Network Nodes}
Network nodes are all endpoints in the network (i.e., every client, relay, and server).
While we could assign their common properties to each of the elements discussed in the remainder of this section, we group them together to reflect their commonality (and to conserve space).

\looseness-1
Some properties, such as control overhead, could arguably be positioned as part of the network itself, but are in this ontology considered part of the network nodes.
The deciding factor was whether the variable could be directly configured or measured as a property of a particular node.
For example, while packet loss requires knowledge of the link between two relays, control overhead can be measured on only one end; therefore, we place the former as a network property and the latter as a property of the network node.
From a more empirical perspective, tools such as Shadow and NetMirage would configure/measure packet loss on the edges of a network graph, while control overhead would be measured using data collected from the node.

\paragraph{Clients}
Clients are the subclass of network nodes that run applications proxied via Tor; they represent both normal Tor clients, as well as onion services.
Client properties include those relating to the Tor application itself,
as well as the application(s) being proxied through it.

\paragraph{Relays}
Relays are the subclass of network nodes that run Tor relays.
As above, relay properties include those of the Tor application, as well as the environment in which it runs.

\paragraph{Servers}
Servers are the subclass of network nodes that represent non-Tor network entities; e.g., web servers and non-Tor clients.
Because they do not run Tor, and will typically be creating requests or responding to requests created elsewhere, they add few properties not already captured above.

\section{Description of Tor Modeling Tools} \label{sec:netmodel:tools:appendix}
Here we provide additional details about the Tor modeling tools
that we described in \S\ref{sec:modeltools}.

\paragraph{TorNetTools}
We wrote a new Tor modeling tool called TorNetTools as a Python package (3,034
LoC) to make our work more accessible. The tool guides researchers through the
experimentation process, including staging and generating network models,
simulating the models, and analyzing and visualizing the results.
The tool implements our modeling algorithms from
\S\ref{sec:tormodel:generate}, producing valid configuration
files that can immediately be used to run Tor experiments in our Tor
experimentation platform as we describe in \S\ref{sec:platforms}.
The tool also implements our statistical analysis methodology from
\S\ref{sec:significance}, including the computation and visualization
of confidence intervals when plotting CDFs. TorNetTools v1.1.0
was released in conjunction with this work.\footnote{https://github.com/shadow/tornettools}

\paragraph{TGen}
We significantly extended and enhanced the TGen traffic
generator~\cite[\S5.1]{tmodel-ccs2018} (6,531 LoC added/modified and
1,411 removed) to support our traffic generation models. Our
improvements include refactoring and generalizing the generator
component so that it can \textit{continuously} generate flows in
addition to streams and packets. We also refactored and extended the
stream component to support deterministic execution of the models,
which allows the client and server sides to synchronize the traffic
generation process by exchanging a seed rather than exchanging the
full models during each stream handshake. Finally, we extended
encoding and validation of Markov models and added support for the
normal, Pareto, and uniform probability distributions (in addition to
the already-supported log-normal and exponential distributions)
to support further exploration of new traffic models.
TGen v1.0.0 was released in conjunction with this
work.\footnote{https://github.com/shadow/tgen}

\paragraph{OnionTrace}
We implemented a new tool called OnionTrace (2,594~LoC) to interact
with a Tor process during an experiment. OnionTrace supports three
modes. In \textit{log} mode, it connects to Tor over Tor's control
port, registers for various asynchronous events (e.g., bytes sent and
received every second), and logs this information to a file for later
analysis. In \textit{record} mode, it extracts information about the
circuits and streams that a Tor process creates over time and stores
it in a \texttt{CSV} file. In \textit{playback} mode, it reads a
previously recorded \texttt{CSV} file and forces Tor to build the same
set of circuits at the same time as indicated in the file. By using
record and playback mode in successive experiments, a researcher can
be sure that the same set of circuits with the same chosen relays are
used across both experiments, improving reproducibility.
OnionTrace v1.0.0 was released in conjunction with this
work.\footnote{https://github.com/shadow/oniontrace}

\newpage
\section{Description of Shadow Improvements} \label{sec:platforms:improve:appendix}
Here we provide additional details about the improvements we made to
the Shadow simulator that we introduced in \S\ref{sec:platforms:improve}.
Our improvements have been merged into Shadow v1.13.2.\footnote{https://github.com/shadow/shadow}

\paragraph{Determinism}
We corrected some bugs in Shadow that caused it to produce
non-deterministic behavior in host IP address assignment and the
processing order of packet events.

\paragraph{Run-time Optimizations}
Through analysis with the performance analysis tool \texttt{perf}, we
identified a bottleneck in Shadow's \texttt{epoll} I/O event
notification facility; our fix reduced run time by 40\% in a small Tor
test network.

\paragraph{Bootstrapping}
We added a new feature that allows configuration of a startup phase
where Shadow will not enforce bandwidth limits or drop packets,
reducing Tor bootstrapping time from 20 to 5 simulated minutes.

\paragraph{Connection Limit}
We associated network connections by 
\texttt{<src\_ip, src\_port, dst\_ip, dst\_port, protocol>}, allowing
each host
to support $2^{16}$ connections to \textit{every} other
host rather than $2^{16}$ connections \textit{globally}.

\paragraph{Network Stack}
We implemented delayed acknowledgments (to reduce the number of empty
ACKs), and explicit duplicate ACKs (to correctly trigger ``3
duplicate ACK'' TCP events).
We updated the TCP auto-tuning algorithm so that it uses the RTT of the peers.
Finally, we corrected a bug by setting packet retransmission timers
and header timestamps when packets are sent rather than when they are
queued to be sent.

\section{Additional Performance Results} \label{app:perf}
We supplement the analysis done in \S\ref{sec:results} with additional plots of
performance metrics measured across our 420 simulations following our
statistical inference methodology from \S\ref{sec:significance}.
Our supplemental results include the transfer times measured by the performance
benchmarking clients: the time to first byte across all transfer sizes are shown
in Figure~\ref{fig:ttfb}, while the time to last byte of 50~KiB and 5~MiB
transfers are shown in Figures~\ref{fig:ttlb:50k} and \ref{fig:ttlb:5m},
respectively. The total Tor network goodput (summed over all relays) is shown in
Figure~\ref{fig:oniontracegput}.

\newcommand\figsizefactor{0.3}

\begin{figure*}[t]
	\centering
	\captionsetup{skip=0pt} 	\begin{subfigure}[b]{\figsizefactor\textwidth}
		\centering
		\captionsetup{skip=0pt} 		\includegraphics[width=1.0\textwidth]{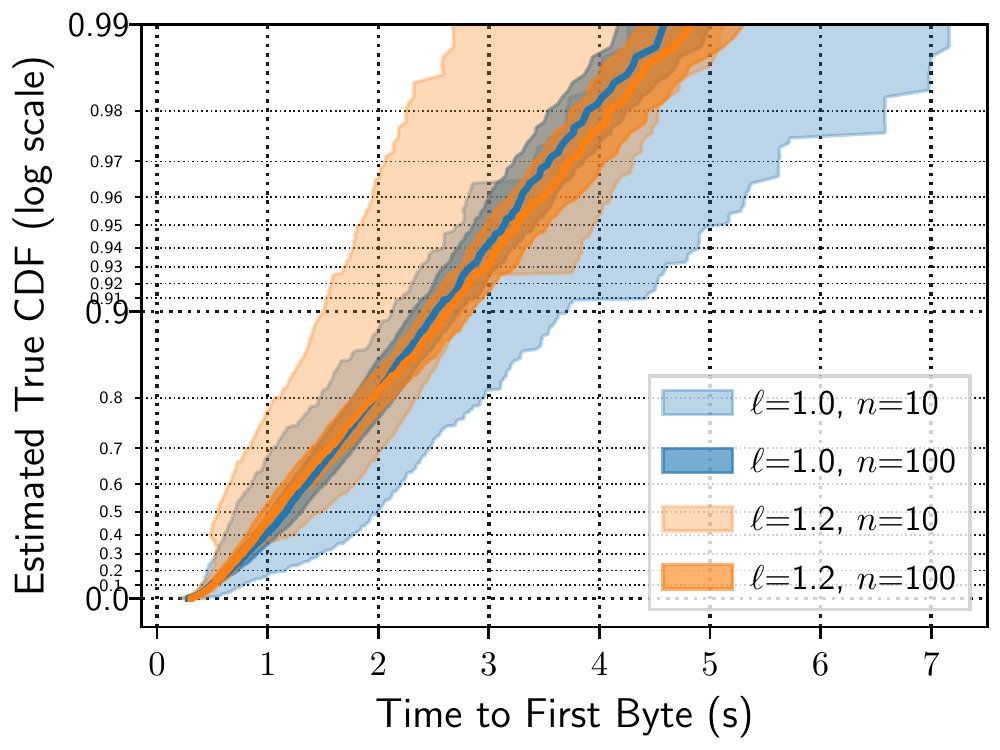}
		\caption{1\% Network Scale ($s = 0.01$)}
	\end{subfigure}
	\begin{subfigure}[b]{\figsizefactor\textwidth}
		\centering
		\captionsetup{skip=0pt} 		\includegraphics[width=1.0\textwidth]{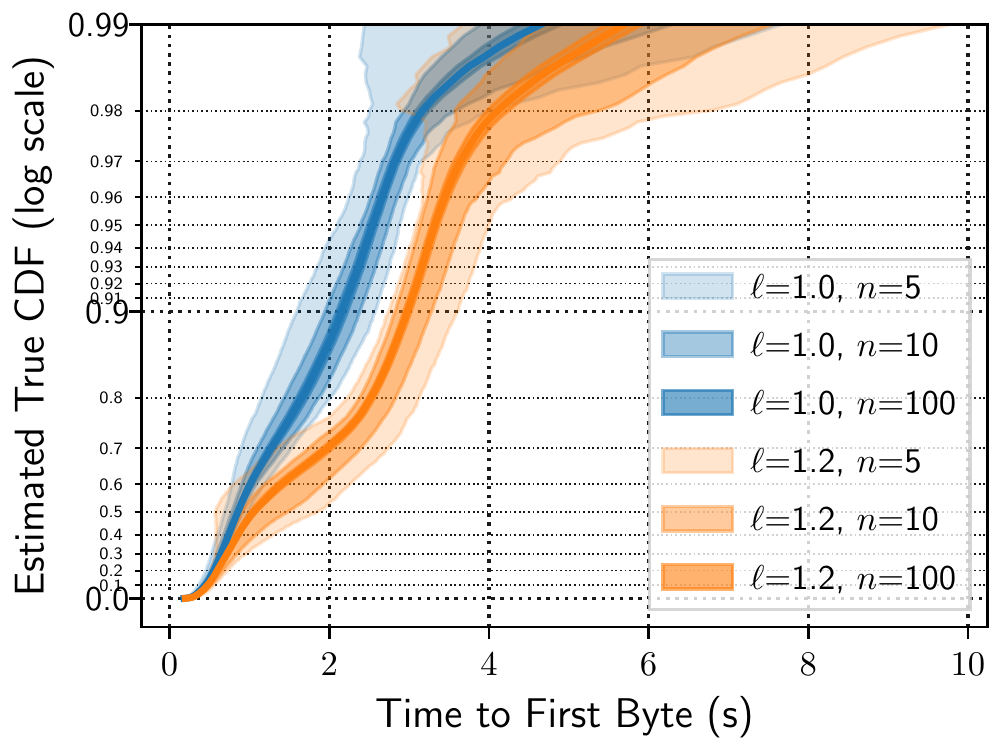}
		\caption{10\% Network Scale ($s = 0.1$)}
	\end{subfigure}
	\begin{subfigure}[b]{\figsizefactor\textwidth}
		\centering
		\captionsetup{skip=0pt} 		\includegraphics[width=1.0\textwidth]{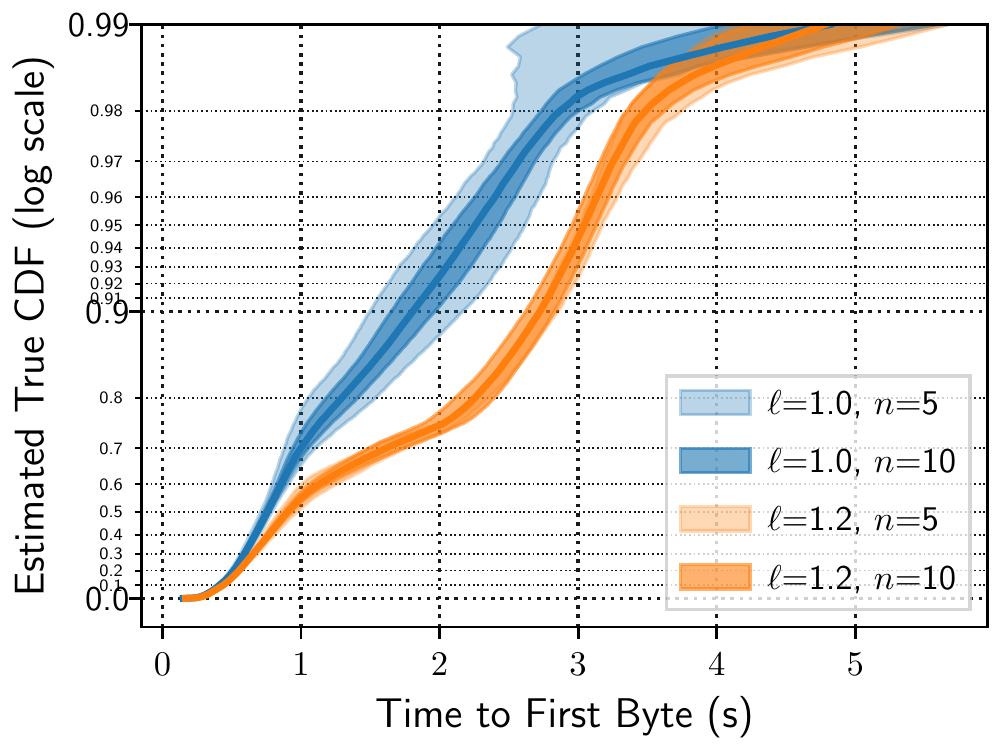}
		\caption{30\% Network Scale ($s = 0.3$)}
	\end{subfigure}
	\caption{
		Time to first byte in seconds of all 50 KiB, 1 MiB, and 5 MiB downloads from performance benchmarking
		clients from experiments with traffic load $\load=1.0$ and $\load=1.2$ in
		networks of various scale $s$. The results from each experiment are aggregated
		from $n$ simulations following \S\ref{sec:significance}, and the CDFs are
		plotted with tail-logarithmic y-axes in order to highlight the long tail of network
		performance.
	}
	\label{fig:ttfb}
	\vspace{-5mm}
\end{figure*}

\begin{figure*}[t]
	\centering
	\captionsetup{skip=0pt} 	\begin{subfigure}[b]{\figsizefactor\textwidth}
		\centering
		\captionsetup{skip=0pt} 		\includegraphics[width=1.0\textwidth]{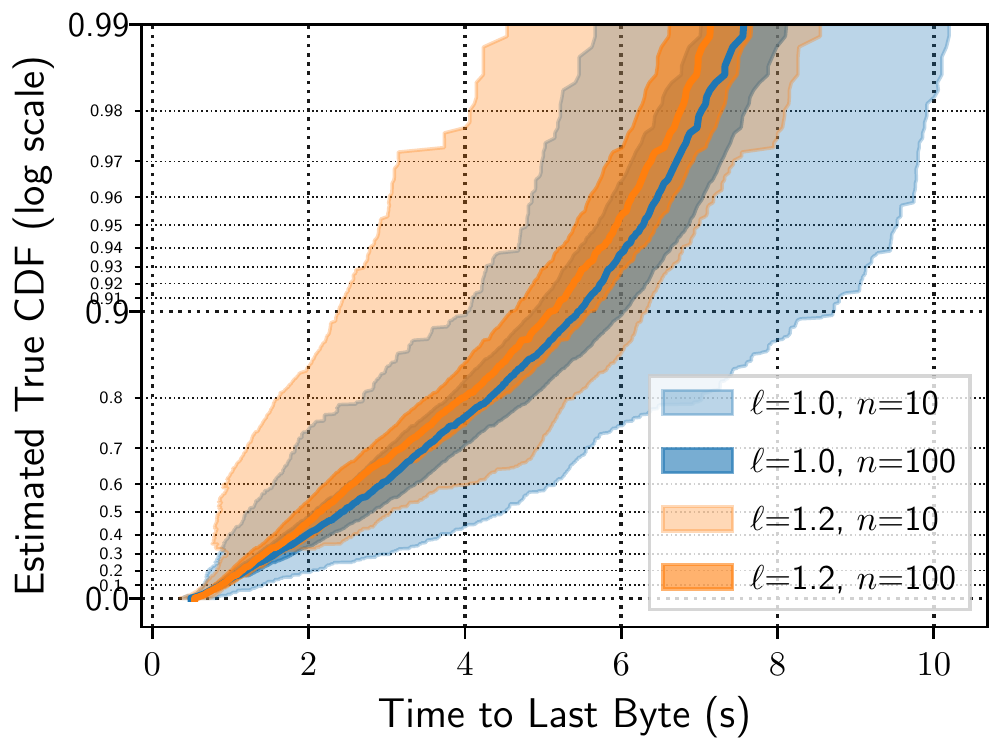}
		\caption{1\% Network Scale ($s = 0.01$)}
	\end{subfigure}
	\begin{subfigure}[b]{\figsizefactor\textwidth}
		\centering
		\captionsetup{skip=0pt} 		\includegraphics[width=1.0\textwidth]{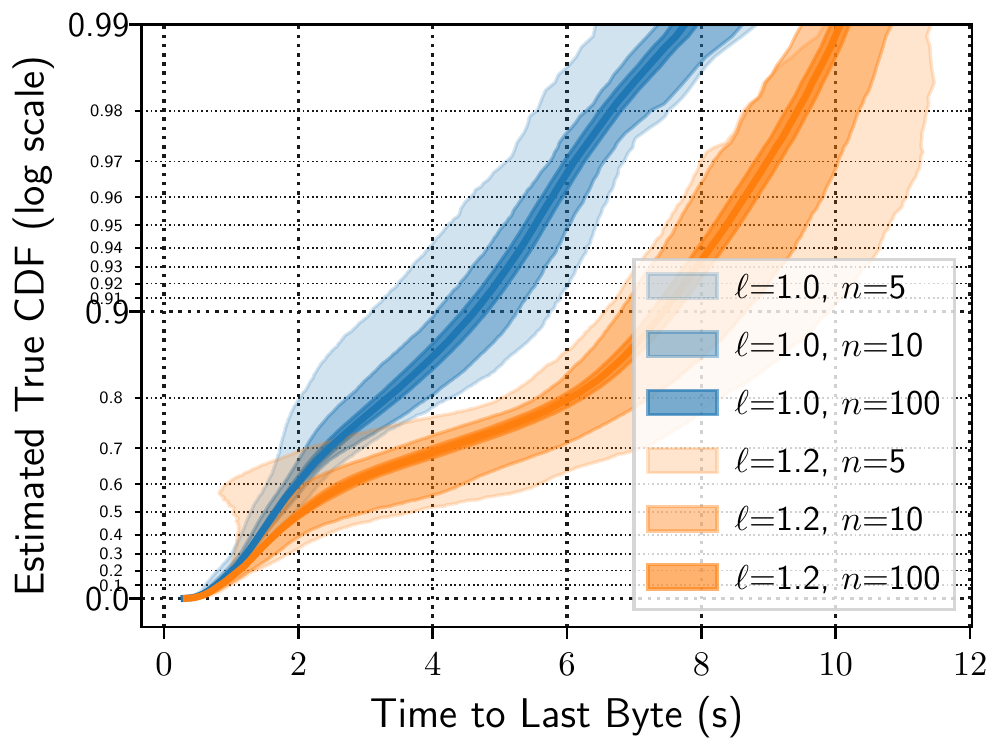}
		\caption{10\% Network Scale ($s = 0.1$)}
	\end{subfigure}
	\begin{subfigure}[b]{\figsizefactor\textwidth}
		\centering
		\captionsetup{skip=0pt} 		\includegraphics[width=1.0\textwidth]{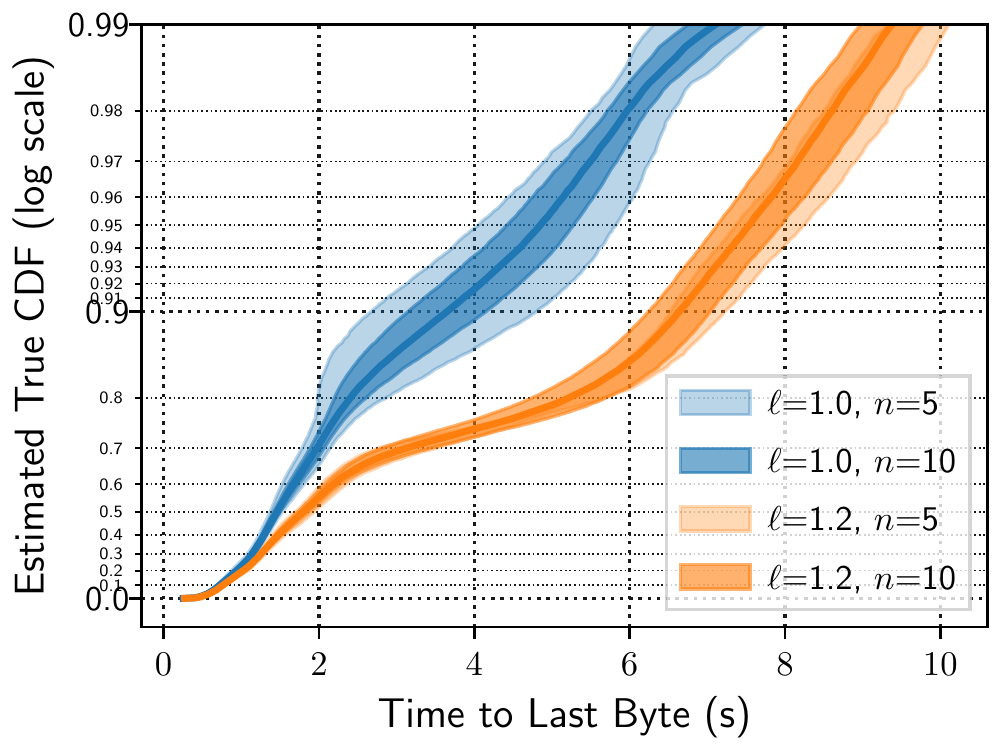}
		\caption{30\% Network Scale ($s = 0.3$)}
	\end{subfigure}
	\caption{
		Time to last byte in seconds of 50~KiB downloads from performance benchmarking
		clients from experiments with traffic load $\load=1.0$ and $\load=1.2$ in
		networks of various scale $s$. The results from each experiment are aggregated
		from $n$ simulations following \S\ref{sec:significance}, and the CDFs are
		plotted with tail-logarithmic y-axes in order to highlight the long tail of network
		performance.
	}
	\label{fig:ttlb:50k}
	\vspace{-5mm}
\end{figure*}

\begin{figure*}[t]
	\centering
	\captionsetup{skip=0pt} 	\begin{subfigure}[b]{\figsizefactor\textwidth}
		\centering
		\captionsetup{skip=0pt} 		\includegraphics[width=1.0\textwidth]{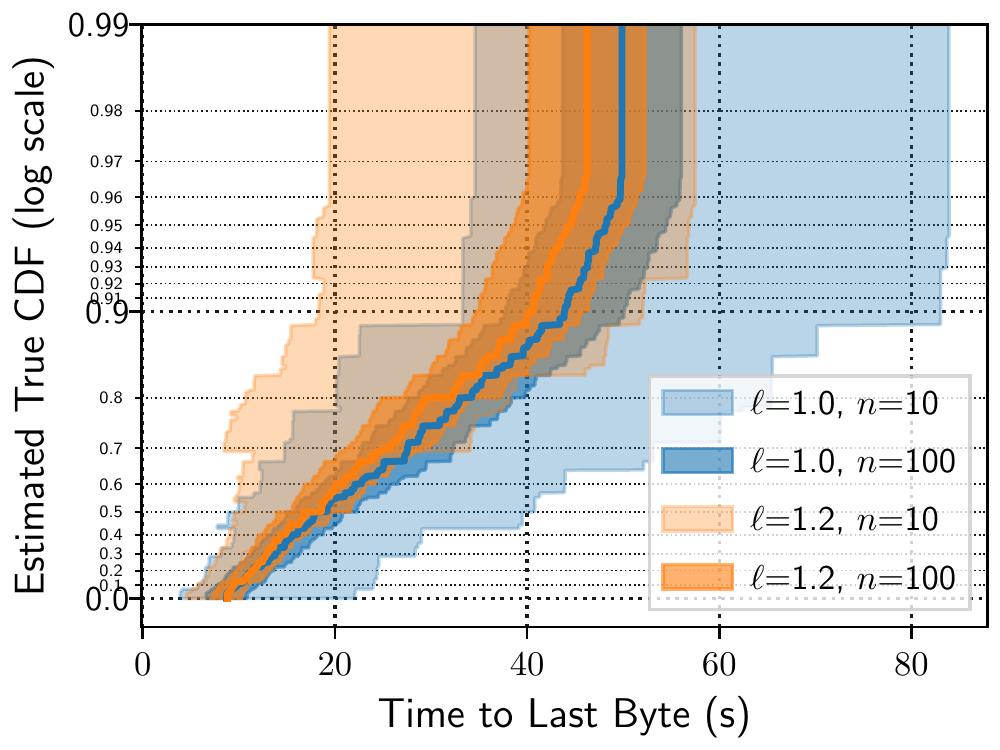}
		\caption{1\% Network Scale ($s = 0.01$)}
	\end{subfigure}
	\begin{subfigure}[b]{\figsizefactor\textwidth}
		\centering
		\captionsetup{skip=0pt} 		\includegraphics[width=1.0\textwidth]{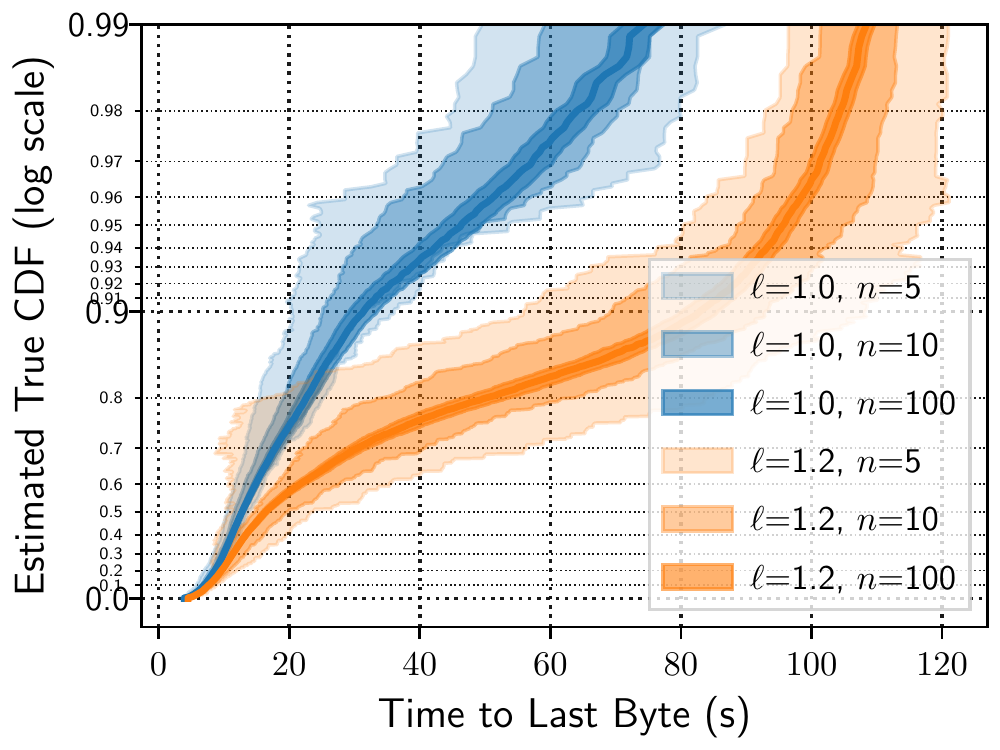}
		\caption{10\% Network Scale ($s = 0.1$)}
	\end{subfigure}
	\begin{subfigure}[b]{\figsizefactor\textwidth}
		\centering
		\captionsetup{skip=0pt} 		\includegraphics[width=1.0\textwidth]{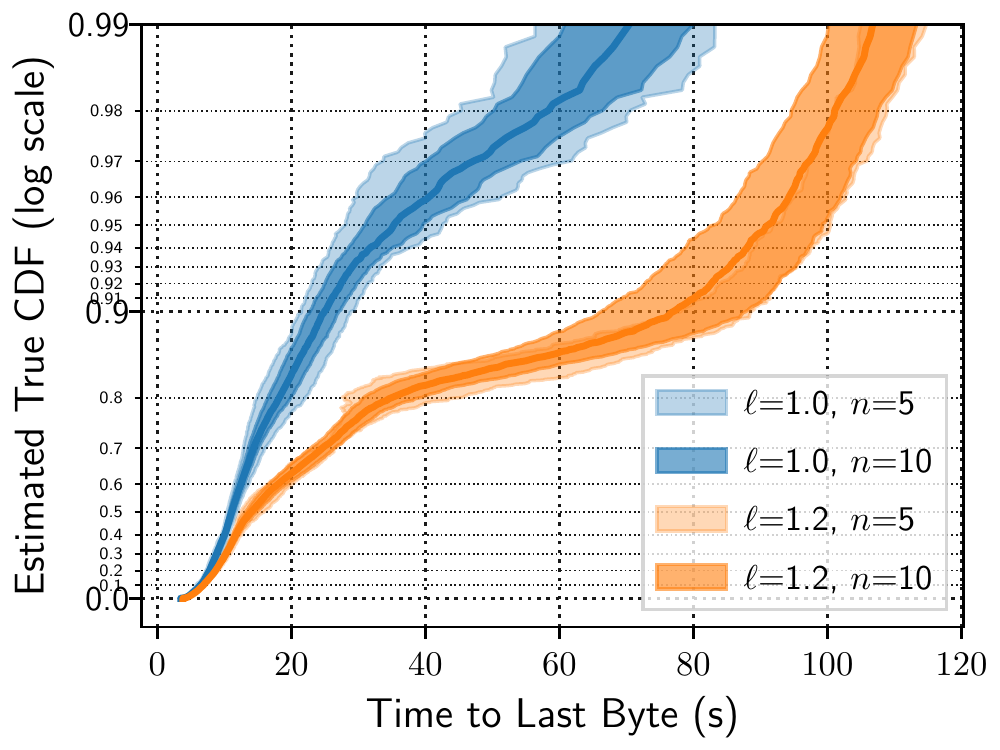}
		\caption{30\% Network Scale ($s = 0.3$)}
	\end{subfigure}
	\caption{
		Time to last byte in seconds of 5~MiB downloads from performance benchmarking
		clients from experiments with traffic load $\load=1.0$ and $\load=1.2$ in
		networks of various scale $s$. The results from each experiment are aggregated
		from $n$ simulations following \S\ref{sec:significance}, and the CDFs are
		plotted with tail-logarithmic y-axes in order to highlight the long tail of network
		performance.
	}
	\label{fig:ttlb:5m}
	\vspace{-5mm}
\end{figure*}

\begin{figure*}[t]
	\centering
	\captionsetup{skip=0pt} 	\begin{subfigure}[b]{\figsizefactor\textwidth}
		\centering
		\captionsetup{skip=0pt} 		\includegraphics[width=1.0\textwidth]{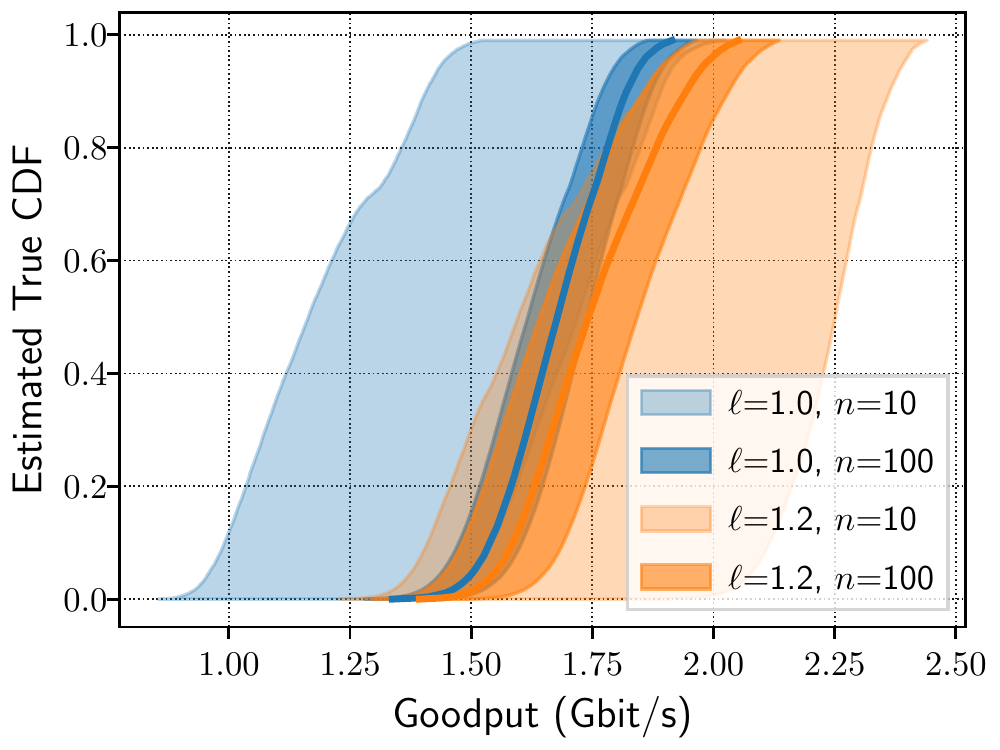}
		\caption{1\% Network Scale ($s = 0.01$)}
	\end{subfigure}
	\begin{subfigure}[b]{\figsizefactor\textwidth}
		\centering
		\captionsetup{skip=0pt} 		\includegraphics[width=1.0\textwidth]{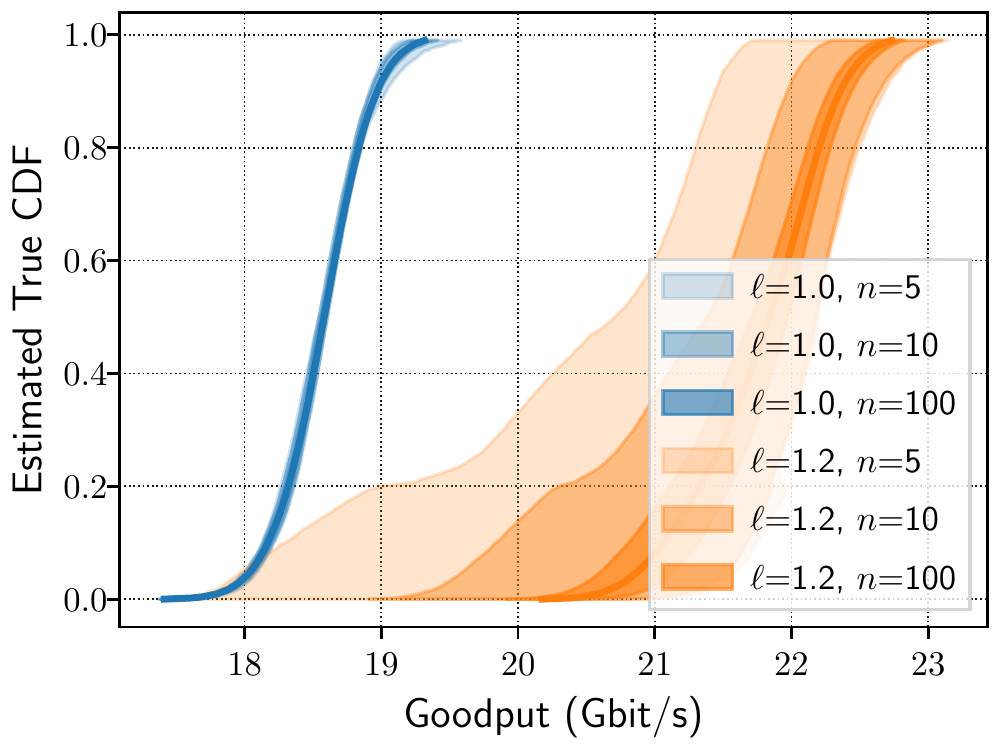}
		\caption{10\% Network Scale ($s = 0.1$)}
	\end{subfigure}
	\begin{subfigure}[b]{\figsizefactor\textwidth}
		\centering
		\captionsetup{skip=0pt} 		\includegraphics[width=1.0\textwidth]{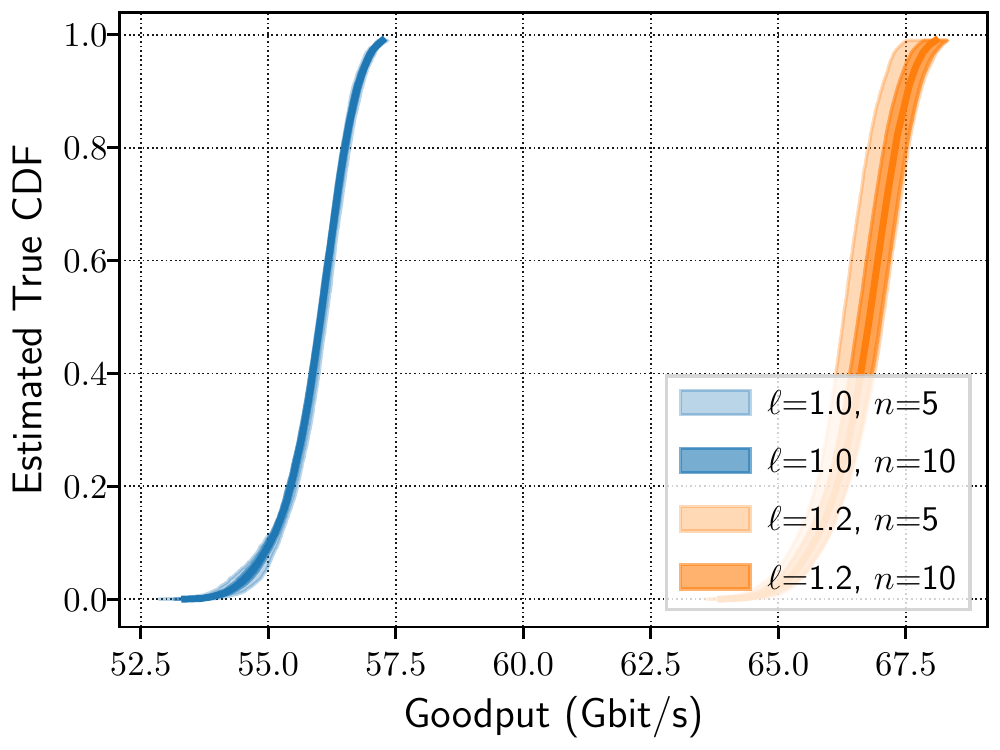}
		\caption{30\% Network Scale ($s = 0.3$)}
	\end{subfigure}
	\caption{
		Tor network goodput in Gbit/s: for each second in the simulation, the sum of
		the goodput from all Tor relays. The results from each experiment are
		aggregated from $n$ simulations following \S\ref{sec:significance}.
	}
	\label{fig:oniontracegput}
	\vspace{-5mm}
\end{figure*}

\end{document}